\documentclass[a4paper,12pt]{article}
\pdfoutput=1
\usepackage{graphicx,subfigure,amsmath,amssymb}
\usepackage{appendix}
\usepackage{color}
\usepackage{cite}

\newlength{\dinwidth}
\newlength{\dinmargin}
\begin{document}
\title{\bf  Probing the effects of new physics in $\bar{B}^* \to P \ell \bar{\nu}_\ell$  decays}
\author{Qin Chang$^{a}$, Jie Zhu$^{a,b}$, Na Wang$^{a}$  and Ru-Min Wang$^{b}$\\
{ $^a$\small Institute of Particle and Nuclear Physics, Henan Normal University, Henan 453007, China}\\
{ $^b$\small Institute of Theoretical Physics, Xinyang Normal University, Henan 464000, China}\\[-0.2cm]
}\date{}
 \maketitle

\begin{abstract}
The significant divergence between the SM predictions and experimental measurements for the ratios, $R_{D^{(*)}}\equiv \mathcal{B}(\bar{B}\to D^{(*)}\tau^- \bar{\nu}_\tau)/ \mathcal{B}(\bar{B}\to D^{(*)} \ell^{\prime-} \bar{\nu}_{\ell^{\prime}})$ with $(\ell^{\prime}=e\,,\mu)$,  implies possible hint of new physics in the flavour sector. In this paper, motivated by the ``$R_{D^{(*)}}$ puzzle" and abundant $B^*$ data samples at high-luminosity heavy-flavor experiments in the future, we try to probe possible effects of  new physics in the semileptonic $\bar{B}^*_{u,d,s} \to P \ell^- \bar{\nu}_\ell$ $(P=D\,,D_s\,,\pi\,,K)$ decays induced by $b \to (u,\,c)\ell^- \bar{\nu}_{\ell}$ transitions in the model-independent vector and scalar scenarios. Using the spaces of NP parameters obtained by fitting to the data of $R_{D}$ and $R_{D^{*}}$, the NP effects on the observables including branching fraction, ratio $R_P^{\ast}$, lepton spin asymmetry and lepton forward-backward asymmetry are studied in detail. We find that the vector type couplings have large effects on the branching fraction and ratio $R_P^{\ast}$. Meanwhile, the scalar type couplings provide significant contributions to all of the observables. The future measurements of these observables in the $\bar{B}^* \to P \ell^- \bar{\nu}_\ell$ decays at the LHCb and Belle-II could provide a way to crosscheck the various NP  solutions to the ``$R_{D^{(*)}}$ puzzle".

\end{abstract}


\newpage
\section{Introduction}
Thanks to the fruitful running of the $B$ factories and Large Hadron Collider~(LHC) in the past years, most of the $B_{u,d}$ mesons decays with branching fractions $\gtrsim$ $\mathcal{O}(10^{-7})$ have been  measured. The rare $B$-meson decays play an important role in testing the standard model~(SM) and probing possible hints of new physics~(NP). Although most of the experimental measurements are in good agreement with the SM predictions, several indirect hints for NP, the tensions or the so-called puzzles, have been observed in the flavour sector.

The semileptonic $\bar{B}\to D^{(*)} \ell \bar{\nu}_{\ell}$ decays are induced by the CKM favored tree-level charged current, and therefore, their physical observables could be rather reliably predicted in the SM and the effects of NP are expected to be tiny. In particular, the ratios defined by $R_{D^{(*)}}\equiv \frac{\mathcal{B}(\bar{B}\to D^{(*)}\tau^- \bar{\nu}_\tau)}{\mathcal{B}(\bar{B}\to D^{(*)} \ell^{\prime-} \bar{\nu}_{\ell^{\prime}})}$~$(\ell^{\prime}=e\,,\mu)$ are independent of the CKM matrix elements, and the hadronic uncertainties canceled to a large extent, thus they could be predicted with a rather high accuracy. However, the BaBar~\cite{Lees:2012xj,Lees:2013uzd}, Belle~\cite{Huschle:2015rga,Sato:2016svk,Abdesselam:2016xqt} and LHCb~\cite{Aaij:2015yra} collaborations have recently observed some anomalies in these ratios. The latest experimental average values for  $R_{D^{(*)}}$ reported by the Heavy Flavor Average Group~(HFAG) are~\cite{Amhis:2016xyh}
\begin{eqnarray}\label{eq:RDHFAG}
R^{\text{Exp}}_{D}=0.403\pm0.040\pm0.024\,,\qquad R^{\text{Exp}}_{D^{*}}=0.310\pm0.015\pm0.008\,,
\end{eqnarray}
which deviate from the SM predictions
\begin{eqnarray}\label{eq:RDSM}
R^{\text{SM}}_{D}=0.300\pm0.008\,\text{\cite{Na:2015kha}},\qquad R^{\text{SM}}_{D^{*}}=0.252\pm0.003\, \text{\cite{Fajfer:2012vx}},
\end{eqnarray}
at the levels of $2.2\sigma$ and $3.4\sigma$ errors, respectively. Moreover, when the correlations between $R_D$ and $R_D^{*}$ are taken into account, the tension would reach up to $3.9\sigma$ level~\cite{Amhis:2016xyh}. Besides, the ratio  $R_{J/\psi}\equiv \frac{\mathcal{B}(B_c\to J/\psi \tau^- \bar{\nu}_\tau)}{\mathcal{B}(B_c\to J/\psi \mu^{-} \bar{\nu}_{\mu})}$ has recently been measured by the LHCb collaboration~\cite{Aaij:2017tyk}, which also shows an excess of about $2\sigma$ from the central value range of the corresponding SM predictions $[0.25,0.28]$. In addition, another mild hint of NP in
the $b \to u \ell \bar{\nu}$ induced $B \to \tau \bar{\nu}$ decay has been observed by the BaBar and Belle Collaborations~\cite{Lees:2012ju,Aubert:2009wt,Adachi:2012mm,Kronenbitter:2015kls}; the deviation is at the level of $1.4 \sigma$~\cite{Ciezarek:2017yzh}.

The large deviations in $R_{D^{(*)}}$ and possible anomalies in the other decay channels mentioned above imply possible hints of NP relevant to the lepton flavor violation~(LFV)~\cite{Ciezarek:2017yzh}. The investigations for these anomalies have been made extensively both within model-independent frameworks~\cite{Freytsis:2015qca,Calibbi:2015kma,Alonso:2015sja,Tanaka:2012nw,
Fajfer:2012jt,Becirevic:2016hea,Bhattacharya:2015ida,Bhattacharya:2014wla,Duraisamy:2014sna,
Hagiwara:2014tsa,Dutta:2013qaa,Duraisamy:2013kcw,Biancofiore:2013ki,Bailey:2012jg,
Becirevic:2012jf,Datta:2012qk,Faller:2011nj,Chen:2005gr,Fan:2014,Fan:2015kna,Alok:2016qyh,
Ivanov:2016qtw}, as well as in some specific NP models where the $b \to c \tau \bar{\nu}_{\tau}$ transition is mediated by leptoquarks\cite{Freytsis:2015qca,Calibbi:2015kma,Deppisch:2016qqd,Dumont:2016xpj,
Dorsner:2016wpm,Sakaki:2014sea,Bauer:2015knc,Fajfer:2015ycq,Sahoo:2015pzk,Barbieri:2015yvd,
Sakaki:2013bfa}, charged Higgses~\cite{Freytsis:2015qca,Celis:2012dk,Celis:2016azn,Li:2017jjs,Cline:2015lqp,Kim:2015zla,
Crivellin:2015hha,Hwang:2015ica,Crivellin:2013wna,Sakaki:2012ft,Nierste:2008qe,Kiers:1997zt,
Tanaka:1994ay,Hou:1992sy}, charged vector bosons~\cite{Freytsis:2015qca,Boucenna:2016wpr,Hati:2015awg}, and sparticles~\cite{Das:2016vkr,Zhu:2016xdg,Wei:2018vmk,Deshpande:2012rr}.

In addition to $B$ mesons, the vector ground states of $b\bar{q}$ system, $B^*$ mesons, with quantum number of $n^{2s+1}L_J=1^3S_1$ and $J^P=1^-$~\cite{Isgur:1991wq,Godfrey:1986wj,Eichten:1993ub,Ebert:1997nk}, also can decay through the $b \to (u,\,c) \ell \bar{\nu}_{\ell}$ transitions at quark-level. Therefore, in principle, the corresponding NP effects might enter into the semileptonic $B^*$ decays as well. The $B^*$ decay occurs mainly  through the electromagnetic process $\bar{B}^* \to \bar{B} \gamma$, and the weak decay modes are very rare. Fortunately, thanks to the rapid development of heavy-flavor experiments instruments and techniques, the $B^*$  weak decays are hopeful to be observed by the running LHC and forthcoming SuperKEK/Belle-II experiments~\cite{Abe:2010gxa,Bediaga:2012py,Aaij:2014jba} in the near future.  For instance, the annual integrated luminosity of Belle-II is expected to reach up to $\sim$ 13 $ab^{-1}$ and the $B^*$ weak decays with branching fractions $> {\cal O}(10^{-9})$ are hopeful to be observed~\cite{Abe:2010gxa,Chang:2015jla,Chang:2015ead}. Moreover, the LHC experiment also will provide a lot of experimental information for $B^*$ weak decays due to the much larger beauty production cross-section of $pp$ collision relative to $e^+e^-$ collision~\cite{Aaij:2010gn}.

Recently, some interesting theoretical studies for the $B^*$ weak decays have been made within the SM in Refs.~\cite{Chang:2015jla,Chang:2015ead,Grinstein:2015aua,Wang:2012hu,Zeynali:2014wya,Bashiry:2014qia,Xu:2015eev,Chang:2016eto,Chang:2016cdi}.
In this paper, motivated by the possible NP explanation for the $R_{D^{(*)}}$ puzzles, the corresponding NP effects on the semileptonic $B^*$ decays will be studied in a model independent way.
In the investigation, the scenarios of vector and scalar NP interactions are studied, respectively; their effects on the branching fraction, differential branching fraction, lepton spin asymmetry, forward-backward asymmetry and ratio $R^{*}_P$~($P=D\,,\pi\,,K$)  of semileptonic $B^*$ decays are explored by using the spaces of various NP couplings obtained through the measured  $R_{D^{(*)}}$.

Our paper is organized as follows. In section 2, after a brief description of the effective Lagrangian for the $b \to (u\,,c) \ell \bar{\nu}_{\ell}$ transitions, the theoretical framework and calculations for the $\bar{B}^*\to P \ell \bar{\nu}_{\ell}$ decays in the presence of various NP couplings are presented. Section 3 is devoted to the numerical results and discussions for the effects of various NP couplings. Finally, we give our conclusions in section 4.

\section{Theoretical framework and calculation}
\subsection{Effective Lagrangian and amplitudes}
We employ the effective field theory approach to compute the amplitudes of $\bar{B}^*\to P \ell \bar{\nu}_{\ell}$ decays in a model independent shceme. The most general effective Lagrangian  at $\mu={\cal O}(m_b)$ for the  $b \to p \ell^- \bar{\nu}_\ell$ ($p=u\,,c$)  transition  can be written as~\cite{Tanaka:2012nw,Becirevic:2016hea,Dorsner:2016wpm,Sakaki:2013bfa}
\begin{eqnarray}\label{eq:Leffall}
\mathcal{L}_{\rm eff}&=&-2\sqrt{2}
G_F\sum_{p=u\,,c}V_{pb}\left\{(1+V_L)\bar{p}_L\gamma^{\mu}b_{L}\bar{\ell}_L\gamma_{\mu}\nu_L
+V_R\bar{p}_R\gamma^{\mu}b_{R}\bar{\ell}_L\gamma_{\mu}\nu_L
+\widetilde{V}_L\bar{p}_L\gamma^{\mu}b_{L}\bar{\ell}_R\gamma_{\mu}\nu_R\right.\nonumber\\
&&+\widetilde{V}_R\bar{p}_R\gamma^{\mu}b_{R}\bar{\ell}_R\gamma_{\mu}\nu_R
+S_L\bar{p}_Rb_{L}\bar{\ell}_R\nu_L
+S_R\bar{p}_Lb_{R}\bar{\ell}_R\nu_L
+\widetilde{S}_L\bar{p}_Rb_{L}\bar{\ell}_L\nu_R
+\widetilde{S}_R\bar{p}_Lb_{R}\bar{\ell}_L\nu_R\nonumber\\
&&\left.+T_L\bar{p}_R\sigma^{\mu\nu}b_{L}\bar{\ell}_R\sigma_{\mu\nu}\nu_L
+\widetilde{T}_L\bar{p}_L\sigma^{\mu\nu}b_{R}\bar{\ell}_L\sigma_{\mu\nu}\nu_R\right\}+\text{h.c.}\,,
\end{eqnarray}
where $G_F$ is the Fermi coupling constant, $V_{pb}$ denotes the CKM matrix elements, $P_{L\,,R}=(1\pm\gamma_5)/2$ is the negative/positive projection operator.
Assuming the neutrinos are left-handed and neglecting the tensor  couplings, the effective Lagrangian can be simplified as
\begin{eqnarray}\label{eq:Leff}
\mathcal{L}_{\rm eff}&=&-
\frac{G_F}{\sqrt{2}}\sum_{p=u\,,c}V_{pb}\left\{
(1+V_L)\bar{p}\gamma_{\mu}(1-\gamma_5)b\bar{\ell}\gamma^{\mu}(1-\gamma_5)\nu
+V_R\bar{p}\gamma_{\mu}(1+\gamma_5)b\bar{\ell}\gamma^{\mu}(1-\gamma_5)\nu\right.\nonumber\\
&&\left.+S_L\bar{p}(1-\gamma_5)b\bar{\ell}(1-\gamma_5)\nu
+S_R\bar{p}(1+\gamma_5)b\bar{\ell}(1-\gamma_5)\nu\right\}+\text{h.c.}\,,
\end{eqnarray}
where, $V_{L,R}$ and $S_{L,R}$ are the effective NP couplings~(Wilson coefficients) defined at $\mu={\cal O}(m_b)$. In the SM, all the NP couplings will be zero.

We use the method of Refs.~\cite{Korner:1987kd,Korner:1989qb,Hagiwara:1989cu,Hagiwara:1989gza,Kadeer:2005aq} to calculate the helicity amplitudes. The square of amplitudes  for the $\bar{B}^* \to P \ell^- \bar{\nu}_\ell$ decay can be written as the product of  leptonic ($L_{\mu\nu}$) and hadronic ($H^{\mu\nu}$) tensors,
\begin{eqnarray}\label{eq:M2}
|{\cal M}(\bar{B}^* \to P \ell^- \bar{\nu}_\ell)|^2=|\langle P \ell^- \bar{\nu}_\ell|\mathcal{L}_{\rm eff}|\bar{B}^*\rangle|^2=\sum_{i,j}L_{\mu\nu}^{ij}H^{ij,\mu\nu}\,,
\end{eqnarray}
where the superscripts $i$ and $j$ refer to four operators in the effective Lagrangian given by Eq.~(\ref{eq:Leff})~\footnote{The tensors related to the scalar and pseudoscalar operators can be understood through the relations given by Eqs.~(\ref{eq:sv}) and (\ref{eq:pa}). }; 
  in the SM, $i=j$ corresponds to the operator $\bar{p}\gamma_{\mu}(1-\gamma_5)b\bar{\ell}\gamma^{\mu}(1-\gamma_5)\nu$. For convenience in writing, these superscripts are omitted below.
Inserting the completeness relation
\begin{eqnarray}
\sum_{m,n}\bar{\epsilon}_{\mu}(m) \bar{\epsilon}_{\nu}^*(n)g_{mn}=g_{\mu\nu}\,,
\end{eqnarray}
The product of $L_{\mu\nu}$ and $H^{\mu\nu}$ can be further expressed as
\begin{eqnarray}\label{eq:M2LI}
L_{\mu\nu}H^{\mu\nu}=\sum_{m,m^{\prime},n,n^{\prime}} L(m,n)H(m^{\prime},n^{\prime})g_{mm^{\prime}}g_{nn^{\prime}}\,.
\end{eqnarray}
Here, $\bar{\epsilon}_{\mu}$ is the polarization vector of the virtual intermediate states, which is $W^{*}$ boson in the SM and  named as $\omega$ in this paper for convenience of expression. The quantities $L(m,n)\equiv L^{\mu\nu}\bar{\epsilon}_{\mu}(m)\bar{\epsilon}^*_{\nu}(n)$ and $H(m,n)\equiv H^{\mu\nu}\bar{\epsilon}^*_{\mu}(m)\bar{\epsilon}_{\nu}(n)$ are Lorentz invariant, and therefore can be evaluated in different reference frames. In the following evaluation, $H(m,n)$ and $L(m,n)$ will be calculated  in the $B^*$-meson rest frame and the  $\ell-\bar{\nu}_\ell$ center-of-mass frame, respectively.

\subsection{Kinematics for $\bar{B}^* \to P \ell^- \bar{\nu}_\ell$ decays}
In the $B^*$-meson rest frame  with daughter $P$-meson moving in the positive $z$-direction, the momenta of particles $B^*$ and $P$ are
\begin{eqnarray}
 p_{B^*}^{\mu}=(m_{B^*},0,0,0)\,,\quad  p_{P}^{\mu}=(E_P,0,0,|\vec{p}|)\,.
\end{eqnarray}
For the four polarization vectors, $\bar{\epsilon}^{\mu}(\lambda_{\omega}=t,0,\pm)$, one can conveniently choose~\cite{Korner:1987kd,Korner:1989qb}
 \begin{eqnarray}\label{eq:polW}
\bar{\epsilon}^{\mu}(t)=\frac{1}{\sqrt{q^2}}(q_0,0,0,-|\vec{p}|)\,,\quad \bar{\epsilon}^{\mu}(0)=\frac{1}{\sqrt{q^2}}(|\vec{p}|,0,0,-q_0)\,,\quad  \bar{\epsilon}^{\mu}(\pm)=\frac{1}{\sqrt{2}}(0,\pm1,-i,0)\,,
\end{eqnarray}
where $q_0=(m_{B^*}^2-m_P^2+q^2)/2m_{B^*}$ and $|\vec{p}|=\lambda^{1/2}(m_{B^*}^2,m_P^2,q^2)/2m_{B^*}$, with $\lambda(a,b,c)\equiv a^2+b^2+c^2-2(ab+bc+ca)$ and $q^2=(p_{B^*}-p_P)^2$ being the momentum transfer squared, are the energy and momentum of the virtual $\omega$. The polarization vectors of the  initial $B^*$-meson can be written as
 \begin{eqnarray}\label{eq:polB}
\epsilon^{\mu}(0)=(0,0,0,1)\,,\quad  \epsilon^{\mu}(\pm)=\frac{1}{\sqrt{2}}(0,\mp1,-i,0)\,.
\end{eqnarray}

In the $\ell-\bar{\nu}_\ell$ center-of-mass frame, the four momenta of lepton and antineutrino pair are given as
 \begin{eqnarray}
 p_\ell^{\mu}=(E_{\ell}, |\vec{p}_{\ell}|\sin\theta,0,|\vec{p}_{\ell}|\cos\theta)\,,\quad  p_{\nu_\ell}^{\mu}=(|\vec{p}_{\ell}|, -|\vec{p}_{\ell}|\sin\theta,0,-|\vec{p}_{\ell}|\cos\theta)\,,
 \end{eqnarray}
where  $E_{\ell}=(q^2+m_{\ell}^2)/2\sqrt{q^2}$, $|\vec{p}_{\ell}|=(q^2-m_{\ell}^2)/2\sqrt{q^2}$, and $\theta$ is the angle between the $P$ and ${\ell}$ three-momenta. In this frame, the polarization vector $\bar{\epsilon}^{\mu}$  takes the form
 \begin{eqnarray}
\bar{\epsilon}^{\mu}(t)=(1,0,0,0)\,,\quad \bar{\epsilon}^{\mu}(0)=(0,0,0,1)\,, \quad \bar{\epsilon}^{\mu}(\pm)=\frac{1}{\sqrt{2}}(0,\mp1,-i,0)\,.
\end{eqnarray}

\subsection{Hadronic helicity amplitudes}
For the $\bar{B}^* \to P \ell^- \bar{\nu}_\ell$ decay, the hadronic helicity amplitudes $H_{\lambda_{B^*} \lambda_{\omega}}^{V_{L,R}}$ and $H_{\lambda_{B^*}\lambda_{\omega}}^{S_{L,R}}$ are defined by
 \begin{eqnarray}\label{eq:hll}
 H_{\lambda_{B^*} \lambda_{\omega}}^{V_L}(q^2)&=&\bar{\epsilon}^{ *\mu}(\lambda_{\omega})\langle P(p_P)|\bar{p}\gamma_{\mu}(1-\gamma_5)b|\bar{B}^*(p_{B^*},\,\lambda_{B^*})\rangle\,, \\
  H_{\lambda_{B^*} \lambda_{\omega}}^{V_R}(q^2)&=&\bar{\epsilon}^{ *\mu}(\lambda_{\omega})\langle P(p_P)|\bar{p}\gamma_{\mu}(1+\gamma_5)b|\bar{B}^*(p_{B^*},\,\lambda_{B^*})\rangle\,,\\
  \label{eq:Hsl}
   H_{\lambda_{B^*}\lambda_{\omega}}^{S_L}(q^2)&=&\langle P(p_P)|\bar{p}(1-\gamma_5)b|\bar{B}^*(p_{B^*},\,\lambda_{B^*})\rangle\,, \\
     \label{eq:Hsr}
    H_{\lambda_{B^*}\lambda_{\omega}}^{S_R}(q^2)&=&\langle P(p_P)|\bar{p}(1+\gamma_5)b|\bar{B}^*(p_{B^*},\,\lambda_{B^*})\rangle\,,
\end{eqnarray}
which describe the decay of three helicity states of $B^*$ meson into a pseudo-scalar $P$ meson and the four helicity states of  virtual $\omega$. It should be noted that $\lambda_{\omega}$ in $H_{\lambda_{B^*}\lambda_{\omega}}^{S_{L\,,R}}(q^2)$, Eqs.~(\ref{eq:Hsl}) and (\ref{eq:Hsr}), should always be equal to $t$.

For $B^*\to P$ transition, the matrix elements of the vector and axial-vector currents can be written in terms of  form factors $V(q^2)$ and $A_{0,1,2}(q^2)$ as
\begin{eqnarray}
\langle P(p_{P})|\bar{p}\gamma_{\mu} b|\bar{B}^*(\epsilon, p_{B^*})\rangle &=&
-\frac{2iV(q^2)}{m_{B^*}+m_{P}}\varepsilon_{\mu\nu\rho\sigma}
\epsilon^{\nu}p_{P}^{\rho}p_{B^*}^{\sigma}\,,\\
\langle P(p_{P})|\bar{p}\gamma_{\mu}\gamma_5 b|\bar{B}^*(\epsilon, p_{B^*})\rangle &=&
2m_{B^*}A_0(q^2)\frac{\epsilon \cdot q}{q^2}q_{\mu}
+(m_{P}+m_{B^*})A_1(q^2)\left(\epsilon_{\mu}-\frac{\epsilon\cdot q}{q^2}q_{\mu}\right)\nonumber \\
&&+A_2(q^2)\frac{\epsilon \cdot q}{m_{P}+m_{B^*}} \left[(p_{B^*}+p_{P})_{\mu}-\frac{m^2_{B^*}-m^2_{P}} {q^2} q_{\mu}\right]\,,
\end{eqnarray}
with the sign convention $\epsilon_{0123}=-1$.
Furthermore, using the equations of motion,
\begin{eqnarray}
i\partial_{\mu}(\bar{p}\gamma^{\mu}b)&=&[m_b(\mu)-m_p(\mu)]\bar{p}b\,,\\
i\partial_{\mu}(\bar{p}\gamma^{\mu}\gamma_5b)&=&-[m_b(\mu)+m_p(\mu)]\bar{p}\gamma_5b\,,
\end{eqnarray}
one can write the  matrix elements of scalar and pseudoscalars currents as
\begin{eqnarray}\label{eq:sv}
\langle P(p_{P})|\bar{p}b|\bar{B}^*(\epsilon, p_{B^*})\rangle
&=&\frac{1}{m_b(\mu)-m_p(\mu)}q_{\mu}\langle P(p_{P})|\bar{p}\gamma^{\mu}b|\bar{B}^*(\epsilon, p_{B^*})\rangle\nonumber\\
&=&0\,,\\
\label{eq:pa}
\langle P(p_{P})|\bar{p}\gamma_5b|\bar{B}^*(\epsilon, p_{B^*})\rangle
&=&-\frac{1}{m_b(\mu)+m_p(\mu)}q_{\mu}\langle P(p_{P})|\bar{p}\gamma^{\mu}\gamma_5b|\bar{B}^*(\epsilon, p_{B^*})\rangle\nonumber\\
&=&-(\epsilon \cdot q)\frac{2m_{B^*}}{m_b(\mu)+m_p(\mu)}A_0(q^2)\,,
\end{eqnarray}
in which, $m_b(\mu)$ and $m_p(\mu)$ are the running quark masses.

Then, by contracting above hadronic matrix elements with the polarization vectors in the $B^*$-meson rest frame, we obtain five non-vanishing helicity amplitudes
\begin{eqnarray}
H_{0t}(q^2)&=&H_{0t}^{V_L}(q^2)=-H_{0t}^{V_R}(q^2)=\frac{2m_{B^*}|\vec{p}|}{\sqrt{q^2}}A_0(q^2),\\
H_{00}(q^2)&=&H_{00}^{V_L}(q^2)=-H_{00}^{V_R}(q^2)\nonumber\\
&=&\frac{1}{2m_{B^*}\sqrt{q^2}}\left[(m_{B^*}+m_{P})(m^2_{B^*}-m^2_{P}+q^2)A_1(q^2)
+ \frac{4m^2_{B^*}|\vec{p}|^2}{m_{B^*}+m_{P}}A_2(q^2) \right],\\ H_{\pm\mp}(q^2)&=&H_{\pm\mp}^{V_L}(q^2)=-H_{\mp\pm}^{V_R}(q^2)=-(m_{B^*}+m_{P})A_1(q^2)\mp\frac{2m_{B^*}|\vec{p}|}{m_{B^*}+m_{P}}V(q^2),\\
\label{eq:h0ts}
H_{0t}^{\prime}(q^2)&=&H_{0t}^{S_L}(q^2)=-H_{0t}^{S_R}(q^2)=-\frac{2m_{B^*}|\vec{p}|}{{m_b(\mu)+m_c(\mu)}}A_0(q^2)\,.
\end{eqnarray}
It is obvious that only the amplitudes with $\lambda_{B^*}=\lambda_{P}-\lambda_{\omega}=-\lambda_{\omega}$ survive.

\subsection{Leptonic helicity amplitudes}
Expanding the leptonic tensor in terms of a complete set of Wigner's $d^J$-functions~\cite{Fajfer:2012vx,Korner:1987kd,Kadeer:2005aq}, $L_{\mu\nu}H^{\mu\nu}$ can be rewritten as a  compact form
\begin{eqnarray}\label{eq:ampd}
L_{\mu\nu}H^{\mu\nu}&=&\frac{1}{8} \sum_{\lambda_{\ell},\lambda_{\bar{\nu}_{\ell}}, \lambda_{\omega},\lambda_{\omega}^{\prime},\, J,\,J^{\prime}}\,(-1)^{J+J^{\prime}}\,h^i_{\lambda_{\ell},\lambda_{\bar{\nu}_{\ell}}}h^{j*}_{\lambda_{\ell},\lambda_{\bar{\nu}_{\ell}}}\,\delta_{\lambda_{B^*}\,,-\lambda_{\omega}}\,\delta_{\lambda_{B^*}\,,-\lambda_{\omega}^{\prime}}\nonumber\\
&&\times\, d^{J}_{\lambda_{\omega},\lambda_{\ell}-\frac{1}{2}}\,d^{J^{\prime}}_{\lambda_{\omega}^{\prime},\lambda_{\ell}-\frac{1}{2}}\,H^i_{\lambda_{B^*}\lambda_{\omega}}\,H^{j*}_{\lambda_{B^*}\lambda_{\omega}^{\prime}}\,,
\end{eqnarray}
in which, $J$ and $J^{\prime}$ run over $1$ and $0$, $\lambda_{\omega}^{(\prime)}$ and $\lambda_{\ell}$ run over their components, and massless right-handed antineutrinos with $\lambda_{\bar{\nu}_{\ell}}=\frac{1}{2}$.
In Eq.~\eqref{eq:ampd}, the $h_{\lambda_{\ell},\lambda_{\bar{\nu}_{\ell}}}^{i,j}$ are the leptonic helicity amplitudes defined as
\begin{eqnarray}
h^{V_{L,R}}_{\lambda_{\ell},\lambda_{\bar{\nu}_{\ell}}}&=&\bar{u}_{\ell}(\lambda_{\ell})\gamma^{\mu}(1-\gamma_5)\nu_{\bar{\nu}}(\frac{1}{2})\bar{\epsilon}_{\mu}(\lambda_{\omega})\,,\\
h^{S_{L,R}}_{\lambda_{\ell},\lambda_{\bar{\nu}_{\ell}}}&=&\bar{u}_{\ell}(\lambda_{\ell})(1-\gamma_5)\nu_{\bar{\nu}}(\frac{1}{2})\,.
\end{eqnarray}
In the  $\ell-\bar{\nu}_\ell$ center-of-mass frame, taking the exact forms of the spinors and polarization vectors, we finally obtain  four nonvanishing contributions
\begin{eqnarray}
|h^{V_{L,R}}_{-\frac{1}{2},\frac{1}{2}}|^2&=&8(q^2-m_\ell^2)\,,\\
|h^{V_{L,R}}_{\frac{1}{2},\frac{1}{2}}|^2&=&8\frac{m_\ell^2}{2q^2}(q^2-m_\ell^2)\,,\\
|h^{S_{L,R}}_{\frac{1}{2},\frac{1}{2}}|^2&=&8\frac{q^2-m_\ell^2}{2}\text{\label{eq:hShS}}\,,\\
|h^{V_{L,R}}_{\frac{1}{2},\frac{1}{2}}|\times|h^{S_{L,R}}_{\frac{1}{2},\frac{1}{2}}|&=&
8\frac{m_{\ell}}{2\sqrt{q^2}}(q^2-m_\ell^2)\text{\label{eq:hVhS}}\,.
\end{eqnarray}

\subsection{ Observables of $\bar{B}^* \to P \ell^- \bar{\nu}_\ell$ Decays}
With the amplitudes obtained in above subsections, we then present the observables considered in our following evaluations. The double differential decay rate of $\bar{B}^* \to P \ell^- \bar{\nu}_\ell$ decay is written as
\begin{eqnarray}
\frac{d\Gamma}{dq^2d\cos\theta}=\frac{G_F^2|V_{pb}|^2}{(2\pi)^3}\,\frac{|\vec{p}|}{8m_{B^*}^2}\,\frac{1}{3}(1-\frac{m_\ell^2}{q^2})|{\cal M}(\bar{B}^* \to P \ell^- \bar{\nu}_\ell)|^2\,,
\end{eqnarray}
where the factor $1/3$ is caused by averaging over the spins of initial state $\bar{B}^*$.
Using the standard convention for $d^J$-function~\cite{Olive:2016xmw}, we finally obtain the double differential decay rates with a  given leptonic helicity state $(\lambda_{\ell}=\pm\frac{1}{2})$, which are
\begin{eqnarray}\label{eq:DdGml}
\frac{d^2\Gamma[\lambda_\ell=-1/2]}{dq^2d\cos\theta}&=&\frac{G_F^2|V_{pb}|^2|\vec{p}|}{256\pi^3m_{B^*}^2}\,\frac{1}{3}\,q^2\,(1-\frac{m_\ell^2}{q^2})^2\,\nonumber\\
&&\times\bigg\{|1+V_L|^2\left[(1-\cos\theta)^2H_{-+}^2+(1+\cos\theta)^2H_{+-}^2+2\sin^2\theta H_{00}^2\right]\,\nonumber\\
&&+|V_R|^2\left[(1-\cos\theta)^2H_{+-}^2+(1+\cos\theta)^2H_{-+}^2+2\sin^2\theta H_{00}^2\right]\,\nonumber\\
&&-4\mathcal{R}e[(1+V_L)V_R^*]\left[(1+\cos\theta^2)H_{+-}H_{-+}+\sin^2\theta H_{00}^2\right]\bigg\}\,,\\
\label{eq:DdGmp}
\frac{d^2\Gamma[\lambda_\ell=1/2]}{dq^2d\cos\theta}&=&\frac{G_F^2|V_{pb}|^2|\vec{p}|}{256\pi^3m_{B^*}^2}\,\frac{1}{3}\,q^2\,(1-\frac{m_\ell^2}{q^2})^2\,\frac{m_\ell^2}{q^2}\nonumber\\
&&\times\bigg\{(|1+V_L|^2+|V_R|^2)\left[\sin^2\theta(H_{-+}^2+H_{+-}^2)+2(H_{0t}-\cos\theta H_{00})^2\right]\nonumber\\
&&-4\mathcal{R}e[(1+V_L)V_R^*]\left[\sin^2\theta H_{-+}H_{+-}+(H_{0t}-\cos\theta H_{00})^2\right]\nonumber\\
&&+4\mathcal{R}e[(1+V_L-V_R)(S^*_L-S^*_R)]\frac{\sqrt{q^2}}{m_{\ell}}\left[H_{0t}^{\prime}(H_{0t}-\cos\theta H_{00})\right]\nonumber\\
&&+2|S_L-S_R|^2\,\frac{q^2}{m_{\ell}^2}\,H_{0t}^{\prime2}\bigg\}\,.
\end{eqnarray}
Using Eqs.~\eqref{eq:DdGml} and \eqref{eq:DdGmp}, ones can get the explicit forms of various observables of $\bar{B}^* \to P \ell^- \bar{\nu}_\ell$ decays as follows:
\begin{itemize}
  \item The differential decay rate
  \begin{eqnarray}\label{eq:SdG}
\frac{d\Gamma}{dq^2}&=&\frac{G_F^2|V_{pb}|^2|\vec{p}|}{96\pi^3m_{B^*}^2}\,\frac{1}{3}\,q^2\,(1-\frac{m_\ell^2}{q^2})^2\nonumber\\
&&\times\bigg\{(|1+V_L|^2+|V_R|^2)[(H_{-+}^2+H_{+-}^2+H_{00}^2)(1+\frac{m_\ell^2}{2\,q^2})+\frac{3m_\ell^2}{2q^2}H_{0t}^2]\nonumber\\
&&-2\mathcal{R}e[(1+V_L)V_R^*][(2H_{-+}H_{+-}+H_{00}^2)(1+\frac{m_\ell^2}{2\,q^2})+\frac{3m_\ell^2}{2q^2}H_{0t}^2]\nonumber\\
&&+3\mathcal{R}e[(1+V_L-V_R)(S_L^*-S_R^*)]H_{0t}^{\prime}H_{0t}\frac{m_{\ell}}{\sqrt{q^2}}+\frac{3}{2}|S_L-S_R|^2H_{0t}^{\prime2}\bigg\}\,.
\end{eqnarray}

  \item The $q^2$ dependent ratio
  \begin{eqnarray}\label{eq:dR}
  R^*_{P}(q^2)\equiv \frac{d\Gamma(\bar{B}^*\to P\tau^-\bar{\nu}_\tau)/dq^2}{d\Gamma(\bar{B}^*\to P\ell^{\prime-}\bar{\nu}_{\ell^{\prime}})/dq^2}\,,
  \end{eqnarray}
where $\ell^{\prime}$ denotes the light lepton.
  \item The lepton spin asymmetry
     \begin{eqnarray}\label{eq:APLamb}
A^P_{\lambda}(q^2)&=&\frac{d\Gamma[\lambda_\ell=-1/2]/dq^2-d\Gamma[\lambda_\ell=1/2]/dq^2}
{d\Gamma[\lambda_\ell=-1/2]/dq^2+d\Gamma[\lambda_\ell=1/2]/dq^2}\,.
\end{eqnarray}

  \item  The forward-backward asymmetry  \begin{eqnarray}\label{eq:APTheta}
A^P_{\theta}(q^2)&=&\frac{\int_{-1}^0d\cos\theta\,( d^2\Gamma/dq^2d\cos\theta)-\int_{0}^1d\cos\theta\,( d^2\Gamma/dq^2d\cos\theta)}{ d\Gamma/dq^2}\,.
\end{eqnarray}
\end{itemize}
The SM results can by obtained from above formulae by taking $V_L=V_R=S_L=S_R=0$.

In the following evaluations, in order to fit the NP spaces, we also need the observables of $\bar{B}\to D^{(*)}\ell^-\bar{\nu}_{\ell}$ decays, which have been fully calculated in the past years. In this paper, we adopt the relevant theoretical formulae given in Ref.~\cite{Sakaki:2013bfa}.

\section{Numerical Results and Discussions}
\subsection{Input Parameters}
Before present our numerical results and analyses,  we would like to clarify the values of input parameters used in the calculation. For the CKM matrix elements, we use~\cite{Charles:2004jd}
\begin{eqnarray}
|V_{cb}|=4.181^{+0.028}_{-0.060}\times10^{-2},\quad |V_{ub}|=3.715^{+0.060}_{-0.060}\times10^{-3}\,.
\end{eqnarray}
For the well-measured Fermi coupling constant $G_F$, the masses of mesons and leptons, and the running masses of quarks at $\mu=m_b$, we take their central values given by PDG~\cite{Olive:2016xmw}. The total decay widths (or  lifetimes) of $B^*$ mesons are essential for estimating the branching fraction, however there is no available experimental data until now. According to the fact that the electromagnetic process $B^*\to B\gamma$ dominates the decays of $B^*$ meson, we take the approximation $\Gamma_{\rm{tot}}(B^*)\simeq \Gamma(B^*\to B\gamma)$; the later has been evaluated within different theoretical models~\cite{Goity:2000dk,Ebert:2002xz,Zhu:1996qy,Aliev:1995wi,Colangelo:1993zq,Choi:2007se,Cheung:2014cka}. In this paper, we adopt the most recent results~\cite{Choi:2007se,Cheung:2014cka}
\begin{eqnarray}
\label{eq:GtotBu}
\Gamma_{\rm{tot}}(B^{*+})&\simeq& \Gamma(B^{*+}\to B^+ \gamma)=(468^{+73}_{-75})\,{\rm eV},\\
 \label{eq:GtotBd}
\Gamma_{\rm{tot}}(B^{*0})&\simeq& \Gamma(B^{*0}\to B^0 \gamma)=(148\pm20)\,{\rm eV},\\
 \label{eq:GtotBs}
\Gamma_{\rm{tot}}(B^{*0}_s)&\simeq& \Gamma(B^{*0}_s\to B^0_s \gamma)=(68\pm17)\,{\rm eV}.
\end{eqnarray}

Then the residual inputs are the transition form factors, which are crucial for evaluating the observables of $\bar{B}^*\to P\ell^-\bar{\nu}_{\ell}$ and $\bar{B}\to D^{(*)}\ell^-\bar{\nu}_{\ell}$  decays. For the $B \to D^{(*)}$ transitions, the scheme of Caprini, Lellouch, and Neubert~(CLN) parametrization~\cite{Caprini:1997mu} is widely used, and the  CLN parameters can be precisely extracted from the well-measured  $\bar{B} \to D^{(*)} \ell^- \bar{\nu}_{\ell}$ decays; numerically, their values read~\cite{Amhis:2016xyh}
\begin{eqnarray}
&&\rho_{D}^{2}=1.128\pm0.033\,,\quad V_1(1)|V_{cb}|=(41.30\pm0.99)\times10^{-3}\,;\\
&&\rho_{D^{*}}^{2}=1.205\pm0.026\,,\quad h_{A_1}(1)|V_{cb}|=(35.38\pm0.43)\times10^{-3},\nonumber\\
&&R_{1}(1)=1.404\pm0.032 \,,\quad R_{2}(1)=0.854\pm0.020  \,.
\end{eqnarray}
However, for the $\bar{B}^{*}_{u,d,s} \to P_{u,d,s}$ transition, there is no experimental data and ready-made theoretical results to use at present. Here, we employ the Bauer-Stech-Wirbel~(BSW) model~\cite{Wirbel:1985ji,Bauer:1988fx} to evaluate the form factors for both $\bar{B}^* \to P$ and  $\bar{B} \to D^{(*)}$ transitions.
Using the inputs $m_u=m_d=0.35\,{\rm GeV}$, $m_s=0.55\,{\rm GeV}$, $m_c=1.7\,{\rm GeV}$, $m_b= 4.9\,{\rm GeV}$ and $\omega=\sqrt{\langle\vec{p}_{\perp}^2\rangle}=0.4\,{\rm GeV}$, we obtain the results at $q^2=0$,
\begin{eqnarray}
&&A^{\bar{B}^{*} \to D}_0(0)=0.71\,,~~~A^{\bar{B}^{*} \to D}_1(0)=0.75\,,~~~
A^{\bar{B}^{*} \to D}_2(0)=0.62\,,~~~V^{\bar{B}^{*} \to D}(0)=0.76\,;\\
&&A^{\bar{B}^{*}_s \to D_s}_0(0)=0.66\,,~~A^{\bar{B}^{*}_s \to D_s}_1(0)=0.69\,,~~
A^{\bar{B}^{*}_s \to D_s}_2(0)=0.59\,,~~V^{\bar{B}^{*}_s \to D_s}(0)=0.72\,;\\
&&A^{\bar{B}^{*} \to \pi}_0(0)=0.34\,,~~~~A^{\bar{B}^{*} \to \pi}_1(0)=0.38\,,~~~~
A^{\bar{B}^{*} \to \pi}_2(0)=0.30\,,~~~~V^{\bar{B}^{*} \to \pi}(0)=0.35\,;\\
&&A^{\bar{B}^{*}_s \to K}_0(0)=0.28\,,~~~A^{\bar{B}^{*}_s \to K}_1(0)=0.29\,,~~~
A^{\bar{B}^{*}_s \to K}_2(0)=0.26\,,~~~V^{\bar{B}^{*}_s \to K}(0)=0.30\,;\\
&&F^{\bar{B} \to D}_0(0)=F^{\bar{B} \to D}_1(0)=0.70\,;\\
&&A^{\bar{B} \to D^{*}}_0(0)=0.63\,,~~~A^{\bar{B} \to D^{*}}_1(0)=0.66\,,~~~
A^{\bar{B} \to D^{*}}_2(0)=0.69\,,~~~V^{\bar{B} \to D^{*}}(0)=0.71\,.
\end{eqnarray}
To be conservative,  $15\%$ uncertainties are assigned to these values in our following evaluation.
Moreover, with the assumption of nearest pole dominance,  the dependences of  form factors on $q^2$  read~\cite{Wirbel:1985ji,Bauer:1988fx}
\begin{eqnarray}
F_0(q^2)&\simeq&\frac{F_0(0)}{1-q^2/m^2_{B_q(0^+)}}\,, \quad
F_1(q^2)\simeq\frac{F_1(0)}{1-q^2/m^2_{B_q(1^-)}}, \nonumber \\
A_0(q^2)&\simeq&\frac{A_0(0)}{1-q^2/m^2_{B_q(0^-)}}\,, \quad
A_1(q^2)\simeq\frac{A_1(0)}{1-q^2/m^2_{B_q(1^+)}}, \nonumber \\
A_2(q^2)&\simeq&\frac{A_2(0)}{1-q^2/m^2_{B_q(1^+)}}\,, \quad
V(q^2)\simeq\frac{V(0)}{1-q^2/m^2_{B_q(1^-)}},
\end{eqnarray}
where $B_q(J^P)$ is the state of $B_q$ with quantum number of $J^P$~($J$ and $P$ are the quantum numbers of total angular momenta and parity, respectively).

With the theoretical formulae and inputs given above, we then proceed to present our numerical results and discussion, which are divided into two scenarios with different simplification for our attention to the types of NP couplings, namely,
\begin{itemize}
\item Scenario I:  taking $S_L=S_R=0$, {\it i.e.}, only considering the NP effects of $V_{L,R}$ couplings\,;
\item Scenario II:  taking $V_L=V_R=0$, {\it i.e.}, only considering the NP effects of $S_{L,R}$ couplings\,.
\end{itemize}
In these two scenarios, we consider all the NP parameters to be real for our analysis. In addition, we assume that only the third generation leptons get corrections from the NP in the $b\to(u,c)\ell\bar{\nu}_{\ell}$ processes and for $\ell=e\,,\mu$ the NP is absent.
In the following discussion, the allowed spaces of NP couplings are obtained by fitting to $R_{D}$ and $R_{D^*}$, Eq.~(\ref{eq:RDHFAG}), with the data varying randomly within their $1\sigma$ error, while the theoretical uncertainties are also considered and obtained by varying the inputs randomly within their ranges specified above.


\subsection{Scenario I: effects of $V_L$ and $V_R$  type couplings}

\begin{figure}[t]
\caption{The allowed spaces of $V_L$ and $V_R$ obtained by fitting to $R_{D}$ and $R_{D^{*}}$. The red and green regions are obtained by using the form factors of CLN parametrization and BSW model, respectively. The right figure shows the minimal result~(solution A) of the four solutions shown in the left figure. }
\begin{center}
\subfigure[]{\includegraphics[scale=0.6]{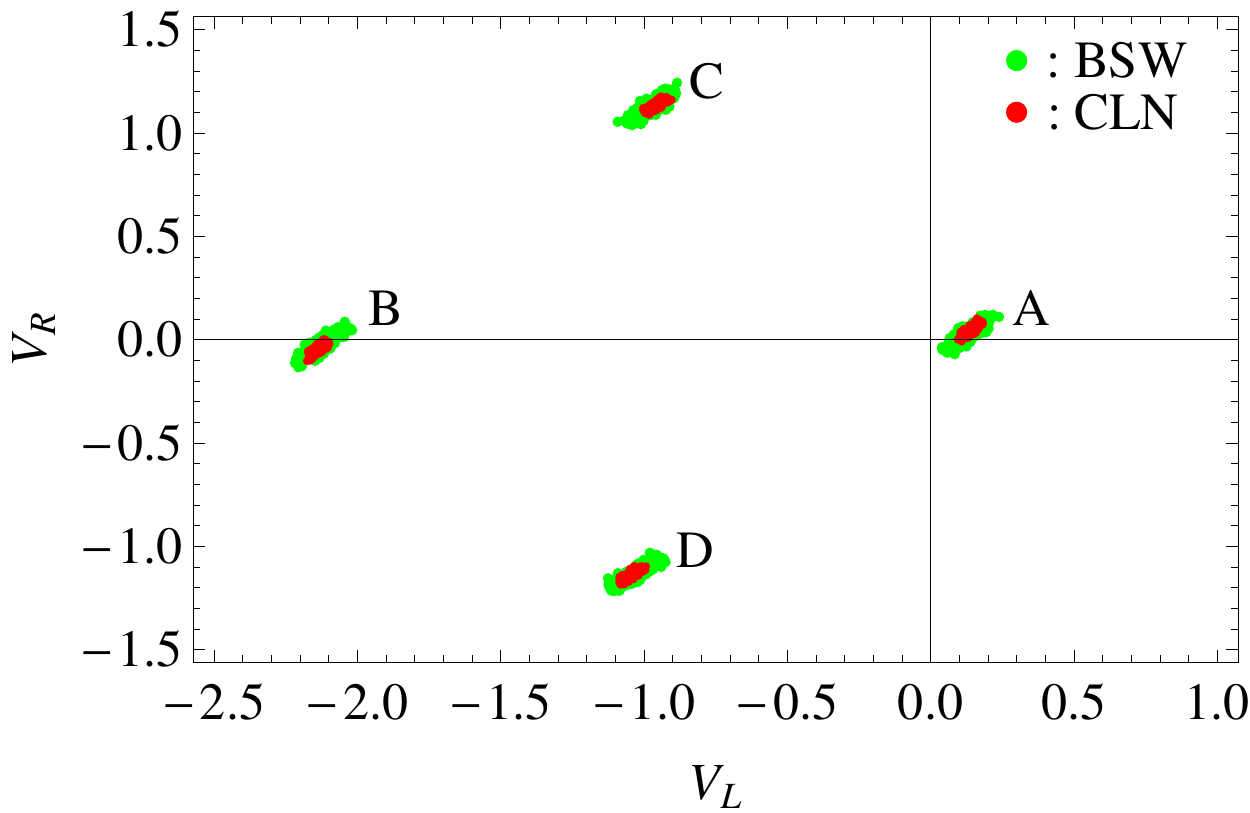}}\qquad\quad
\subfigure[]{\includegraphics[scale=0.6]{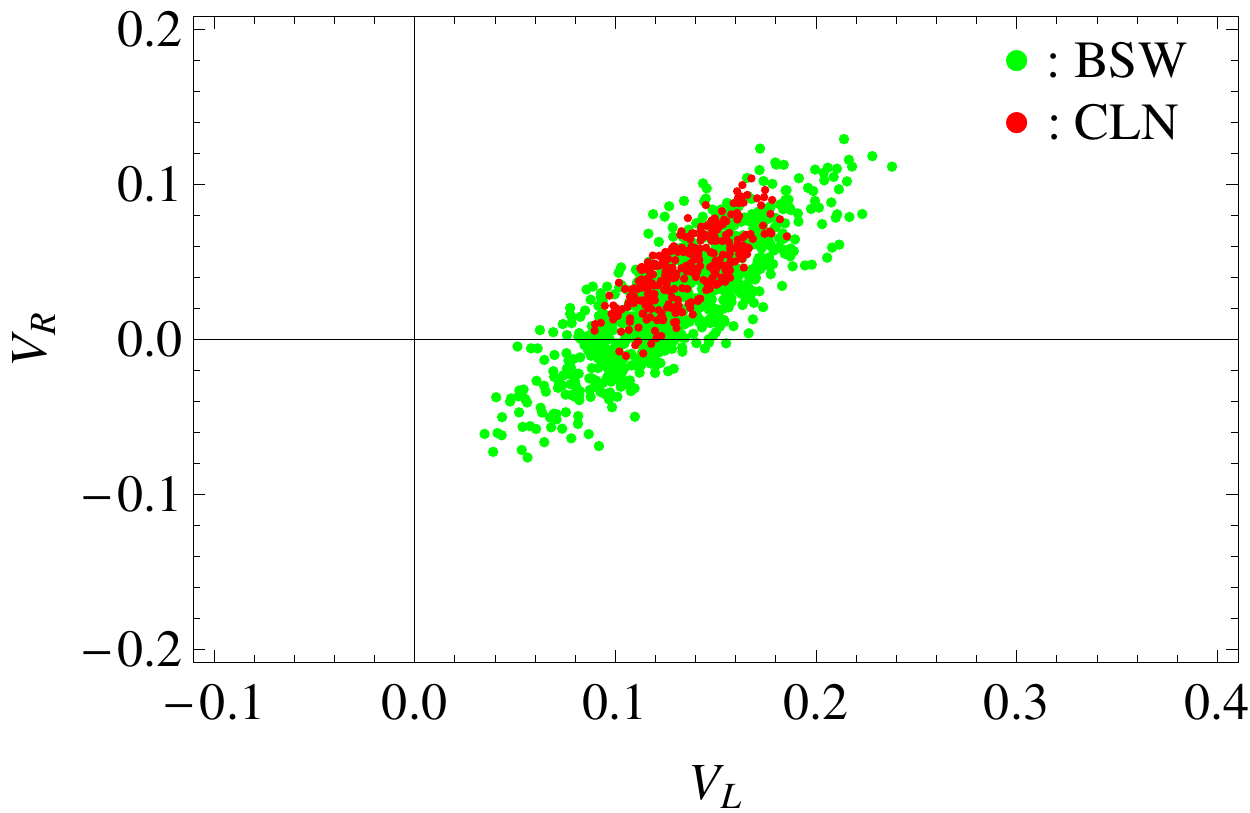}}
\end{center}
\label{fig:VLVR}
\end{figure}

\begin{table}[t]
\caption{The theoretical predictions for the branching fractions of  $ \bar{B}^* \to P \tau^- \bar{\nu}_\tau$ decays  and $R^*_P$ within the SM and the two scenarios. The first error is caused by the uncertainties of form factors, CKM factors and $\Gamma_{tot}(B^*)$; and the second error given in the last two columns is caused by the NP couplings.}
\begin{center}
\begin{tabular}{lccc}
\hline\hline
Obs.                      &SM Prediction&Scenario I&Scenario II\\\hline
$\mathcal{B}(B^{*-} \to D^0\tau^- \bar{\nu}_{\tau})$
  &$0.87^{+0.46}_{-0.32}\times10^{-8}$&$1.04^{+0.54}_{-0.38}$$^{+0.06}_{-0.05}\times10^{-8}$&$1.00^{+0.51}_{-0.36}$$^{+0.03}_{-0.04}\times10^{-8}$\\
$\mathcal{B}(\bar{B}^{*0} \to D^+ \tau^- \bar{\nu}_{\tau})$
  &$2.74^{+1.29}_{-0.94}\times10^{-8}$&$3.27^{+1.66}_{-1.14}$$^{+0.19}_{-0.15}\times10^{-8}$&$3.13^{+1.52}_{-1.13}$$^{+0.10}_{-0.11}\times10^{-8}$\\
$\mathcal{B}(\bar{B}^{*0}_s \to D^+_s \tau^- \bar{\nu}_{\tau})$
  &$5.13^{+3.67}_{-2.13}\times10^{-7}$&$6.13^{+4.51}_{-2.48}$$^{+0.35}_{-0.28}\times10^{-7}$&$5.89^{+3.93}_{-2.39}$$^{+0.20}_{-0.22}\times10^{-7}$\\\hline
$\mathcal{B}(B^{*-} \to \pi^0\tau^- \bar{\nu}_{\tau})$
  &$1.42^{+0.79}_{-0.50}\times10^{-10}$&$1.71^{+0.91}_{-0.63}$$^{+0.09}_{-0.07}\times10^{-10}$&$1.74^{+0.94}_{-0.62}$$^{+0.10}_{-0.10}\times10^{-10}$\\
$\mathcal{B}(\bar{B}^{*0} \to \pi^+ \tau^- \bar{\nu}_{\tau})$
  &$0.99^{+0.38}_{-0.41}\times10^{-9}$&$1.08^{+0.55}_{-0.37}$$^{+0.06}_{-0.05}\times10^{-9}$&$1.09^{+0.52}_{-0.39}$$^{+0.06}_{-0.06}\times10^{-9}$\\
$\mathcal{B}(\bar{B}^{*0}_s \to K^+ \tau^- \bar{\nu}_{\tau})$
&$0.95^{+0.65}_{-0.40}\times10^{-9}$&$1.14^{+0.78}_{-0.46}$$^{+0.06}_{-0.05}\times10^{-9}$&$1.20^{+0.87}_{-0.47}$$^{+0.08}_{-0.08}\times10^{-9}$\\\hline
  $R^*_{D}$&$0.298^{+0.012}_{-0.010}$&$0.355^{+0.015}_{-0.011}$$^{+0.020}_{-0.016}$&$0.341^{+0.048}_{-0.026}$$^{+0.011}_{-0.012}$\\
  $R^*_{\pi}$&$0.677^{+0.013}_{-0.014}$&$0.816^{+0.017}_{-0.012}$$^{+0.044}_{-0.035}$&$0.827^{+0.126}_{-0.073}$$^{+0.046}_{-0.048}$\\
  $R^*_{K}$&$0.638^{+0.017}_{-0.015}$&$0.770^{+0.021}_{-0.015}$$^{+0.042}_{-0.034}$&$0.810^{+0.144}_{-0.084}$$^{+0.052}_{-0.054}$\\\hline
\hline
\end{tabular}
\end{center}
\label{tab:Bstar2P}
\end{table}
\begin{figure}[t]
\caption{The $q^2$-dependence of the differential observables $d\Gamma/dq^2$, $R^*_P$, $A_{\lambda}^P$ and $A_{\theta}^P$ for $B^{*-} \to D^0 \tau^- \bar{\nu}_{\tau}$ and $\pi^0 \tau^- \bar{\nu}_{\tau}$ decays within the SM and scenario I.}
\begin{center}
\subfigure[]{\includegraphics[scale=0.45]{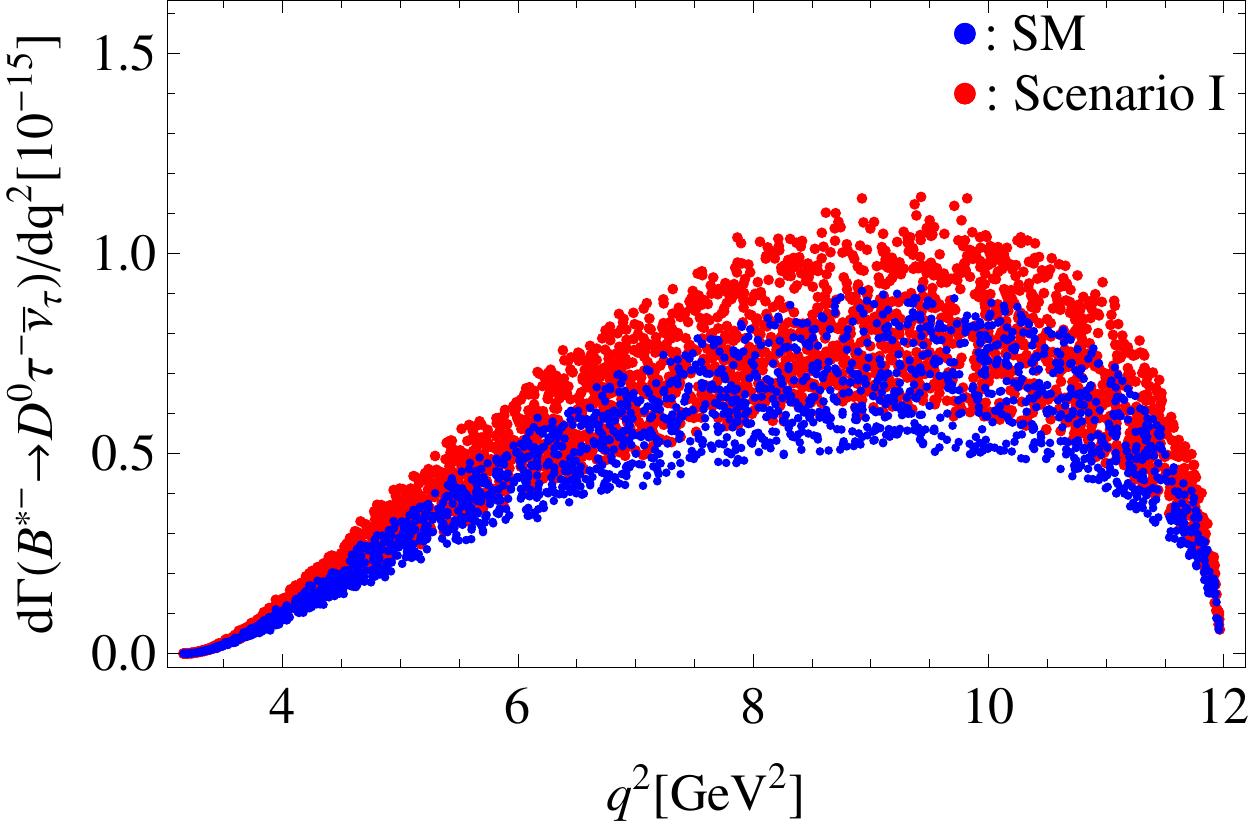}}\qquad
\subfigure[]{\includegraphics[scale=0.45]{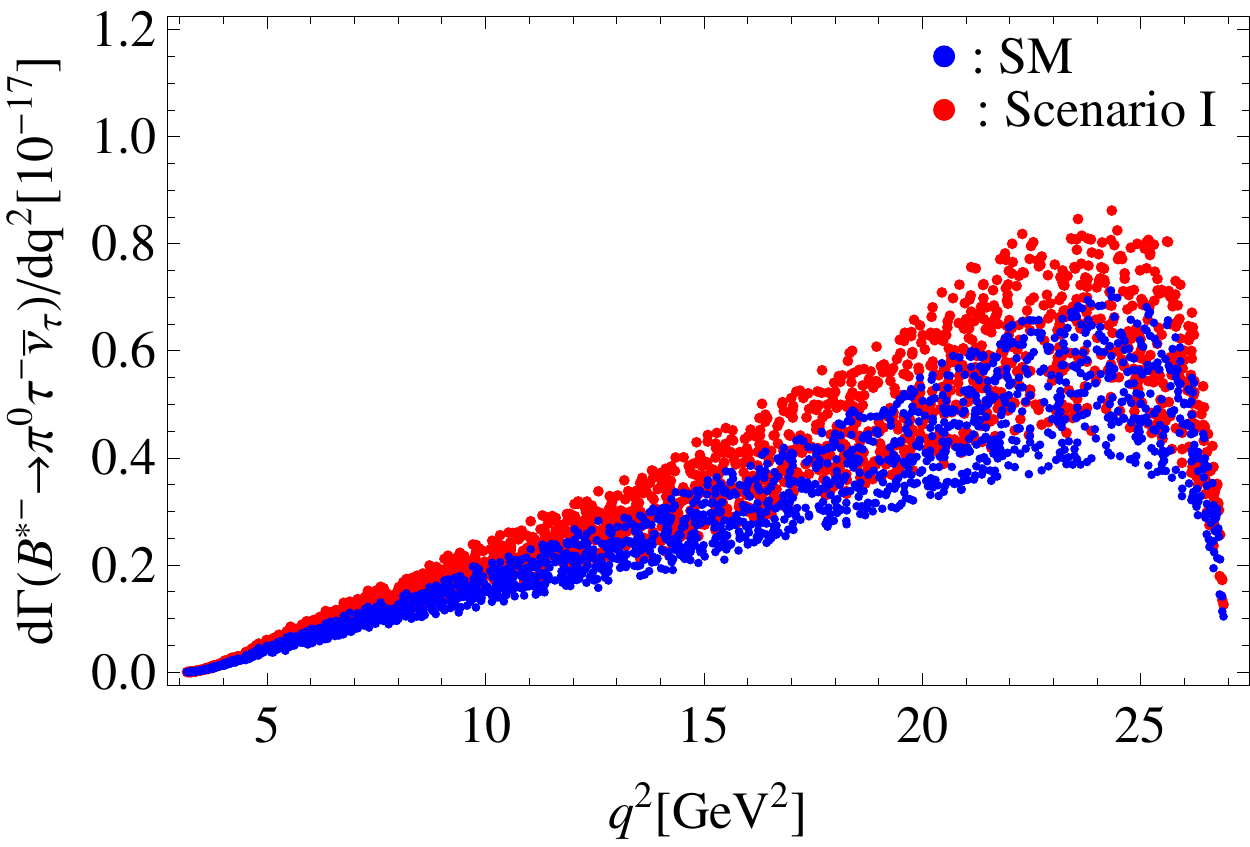}}\\
\subfigure[]{\includegraphics[scale=0.45]{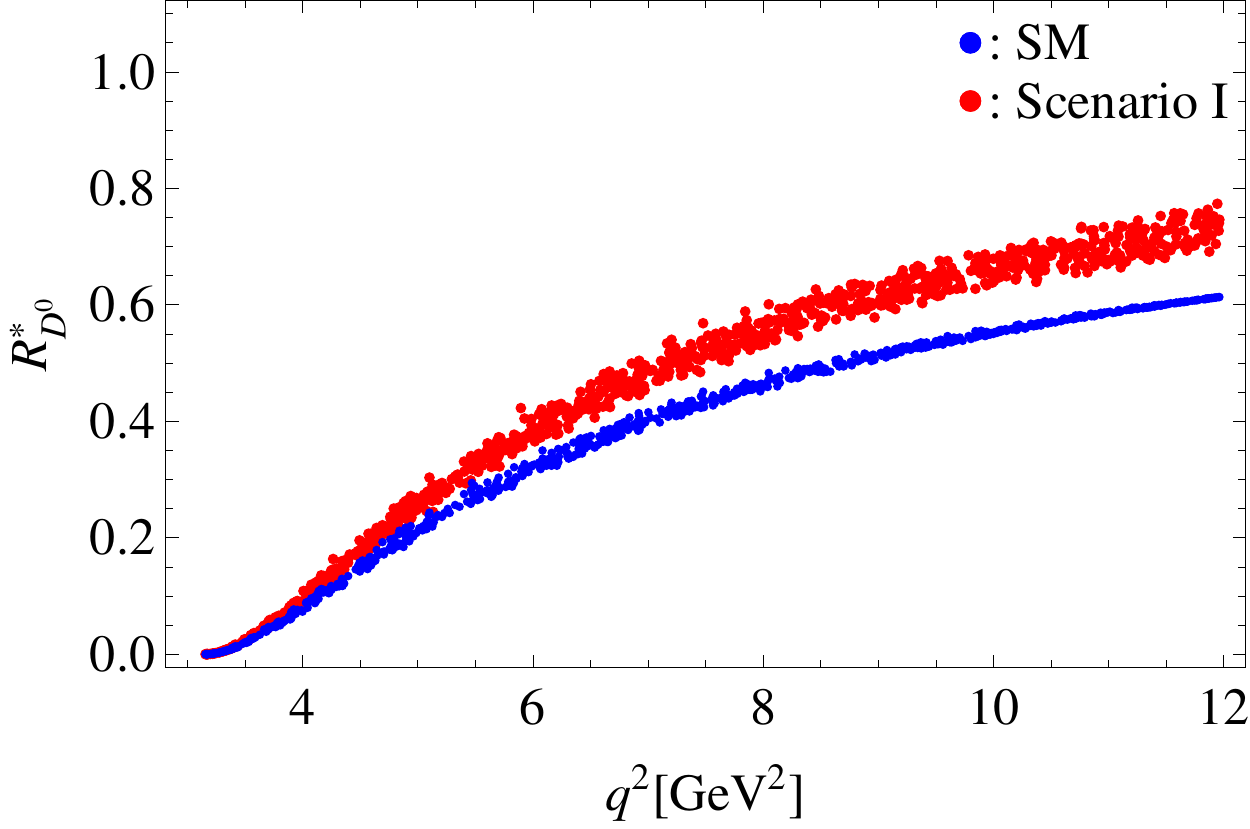}}\qquad
\subfigure[]{\includegraphics[scale=0.45]{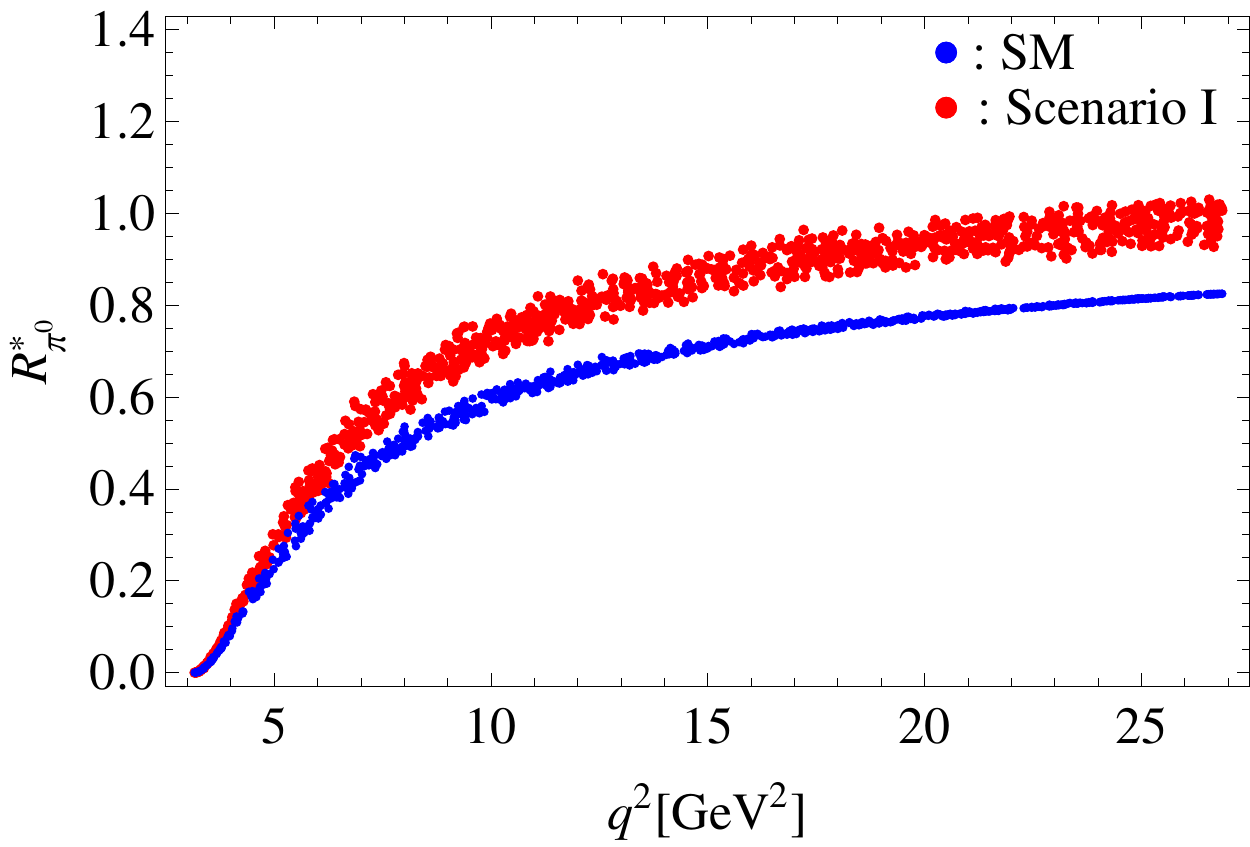}}\\
\subfigure[]{\includegraphics[scale=0.45]{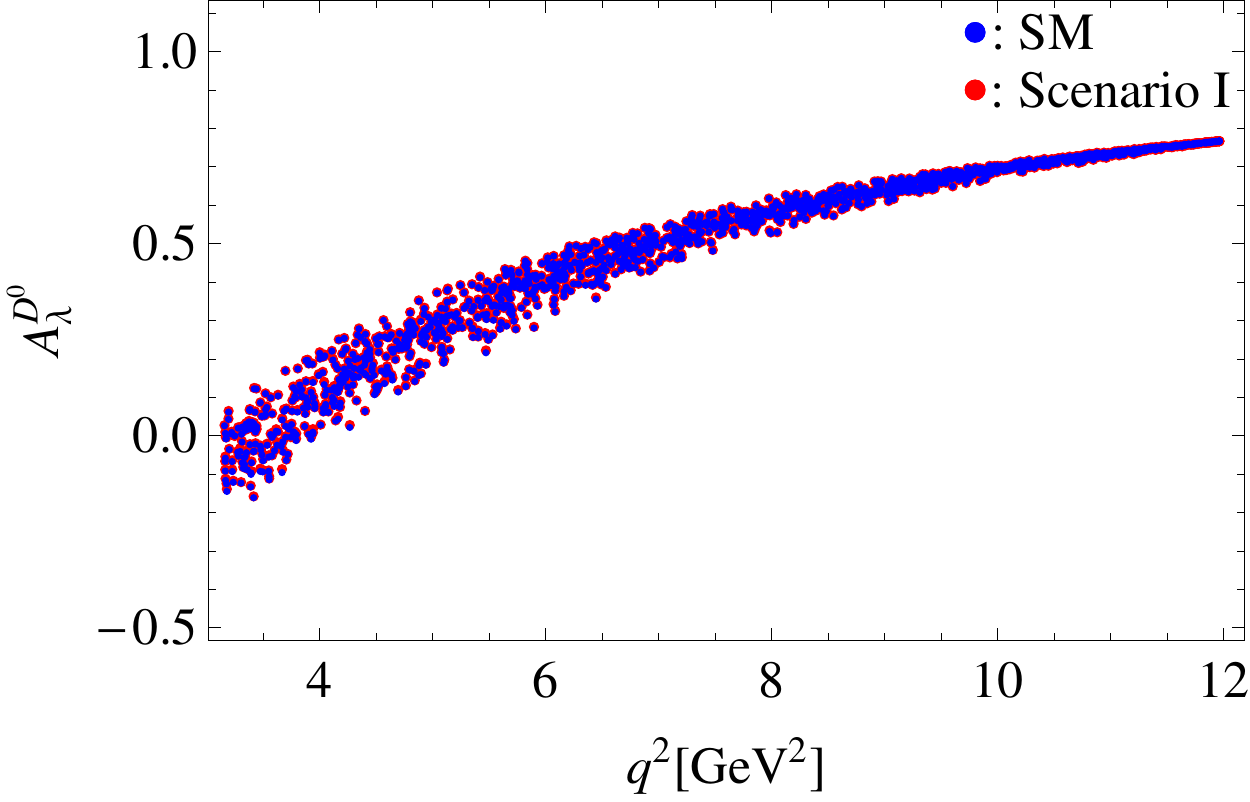}}\qquad
\subfigure[]{\includegraphics[scale=0.45]{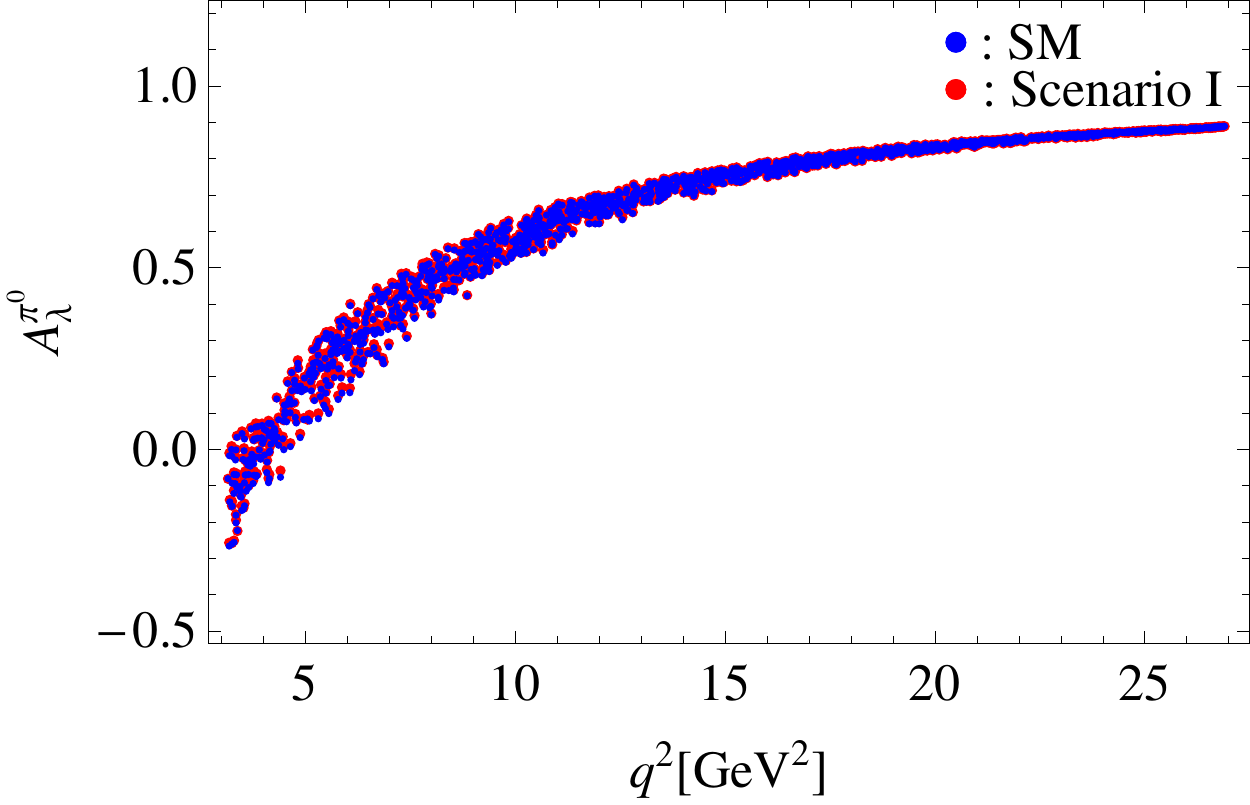}}\\
\subfigure[]{\includegraphics[scale=0.45]{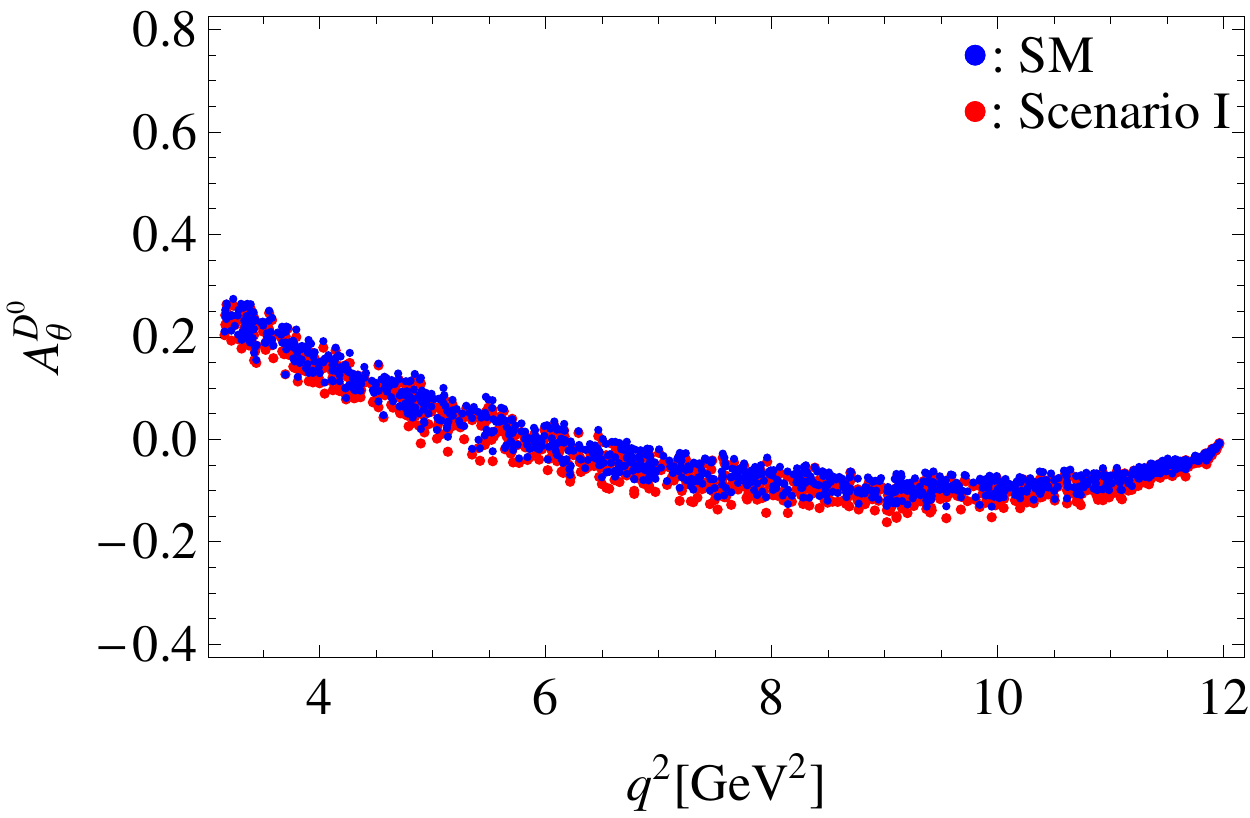}}\qquad
\subfigure[]{\includegraphics[scale=0.45]{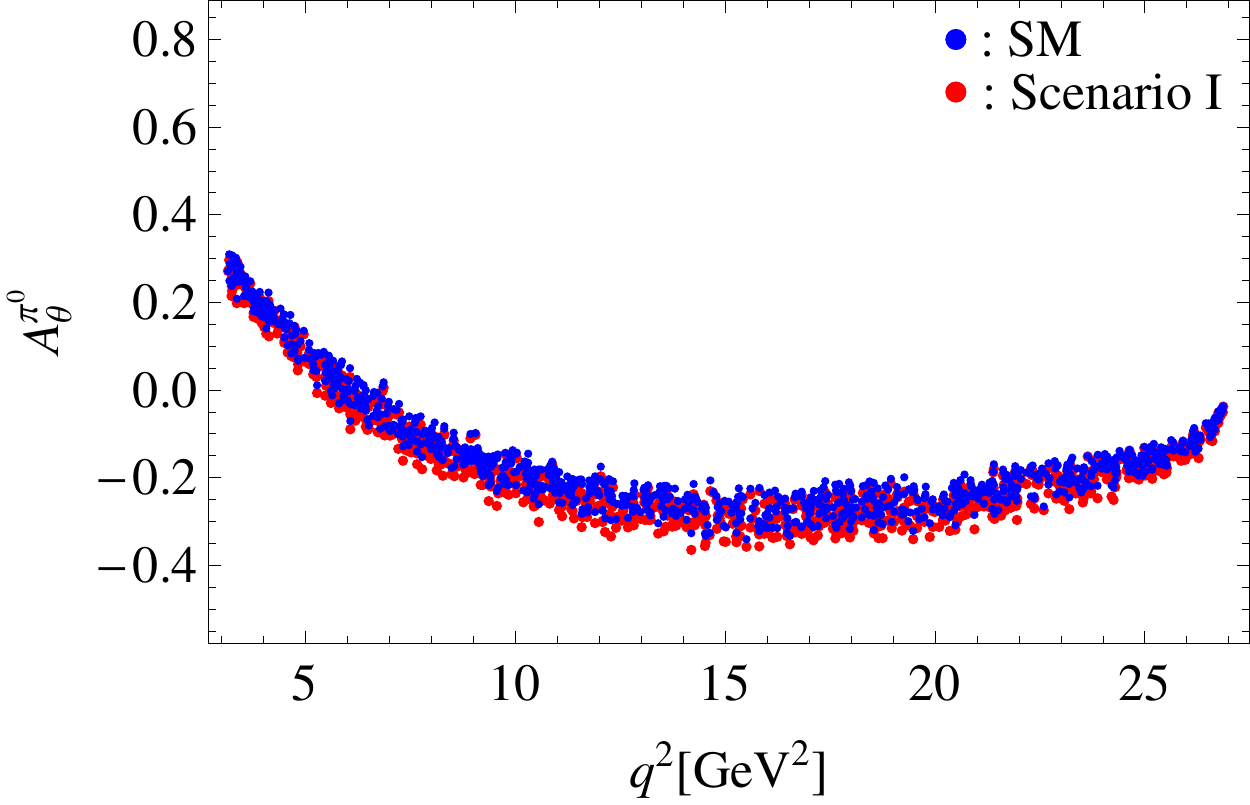}}
\end{center}
\label{fig:Bstar2PCVob1}
\end{figure}

In this subsection, we vary couplings $V_L$ and $V_R$ while keeping all other NP couplings to zero. Under the constraints from the data of  $R_D$ and $R_D^{*}$,
the allowed spaces of new physics parameters, $V_L$ and $V_R$, are shown in the Fig.~\ref{fig:VLVR}. In the fit,  the $B \to D^{(*)}$ form factors based on CLN parametrization and BSW model are used, respectively; it can be seen from Fig.~\ref{fig:VLVR} that their corresponding fitting results are in consistence with each other, but the constraint with the former is much stronger due to the relatively small theoretical error. Therefore, in the following evaluations and discussions, the results obtained by using  CLN parametrization  are used. In addition, our fitting result Fig.~\ref{fig:VLVR} agrees well with the  ones obtained in the previous works, for instance, Refs.~\cite{Dutta:2013qaa,Datta:2012qk}.

From Fig.~\ref{fig:VLVR}, we find that: (i) the allowed spaces of $(V_L,\,V_R)$ are bounded into four separate regions, namely solutions A-D.   (ii) Except for the solution A, the others solutions are all far from the zero point $(0,0)$, and result in very large NP contributions.  Taking the solution C~(D) as an example,  the SM contribution is completely canceled out by the NP contribution related to $V_L$,  and the $V_R$ coupling presents sizable positive~(negative) NP contribution to fit data. The situation of solution B is similar, but only $V_L$ coupling presents sizable NP contribution. Numerically, one can easily conclude that  the NP contributions of solutions B-D are about two times larger than the SM, which seriously exceeds our general expectation that the amplitudes should be dominated by the SM and the NP only presents minor corrections. In this point of view, the minimal solution~(solution A) is much favored  than the solutions B-D.  So, in our following discussions, we  pay attention only to the solution A, which is replotted in Fig.~\ref{fig:VLVR}(b) and numerical result is
 \begin{eqnarray}\label{eq:s1}
V_L=0.14^{+0.06}_{-0.06}\,,\qquad V_R=0.05^{+0.06}_{-0.07}\,.\qquad {\rm solution~A}
\end{eqnarray}

Using the values of NP couplings given by Eq.~(\ref{eq:s1}), we then present our theoretical predictions for $\mathcal{B}(\bar{B}^*\to P \tau^- \bar{\nu}_{\tau})$ and $q^2$-integrated $R^*_P$ in Table~\ref{tab:Bstar2P}, in which the SM results are also listed for comparison.  The $q^2$-dependence of differential observables  $d\Gamma/dq^2$, $R^*_P$, $A_{\lambda}^P$ and $A_{\theta}^P$ for $B^{*-} \to D^0 \tau^- \bar{\nu}_{\tau}$ and $\pi^0 \tau^- \bar{\nu}_{\tau}$ decays are shown in  Fig.~\ref{fig:Bstar2PCVob1}; the case of $\bar{B}^{*0} \to D^+ \tau^- \bar{\nu}_{\tau}$ and $\bar{B}^{*0}_s \to D^+_s \tau^- \bar{\nu}_{\tau}$ ($\bar{B}^{*0} \to \pi^+ \tau^- \bar{\nu}_{\tau}$ and $\bar{B}^{*0}_s \to K^+ \tau^- \bar{\nu}_{\tau}$) are similar to the one of $B^{*-} \to D^0 \tau^- \bar{\nu}_{\tau}$~($B^{*-} \to \pi^0 \tau^- \bar{\nu}_{\tau}$) decay, and not shown here.
The following are some discussions and comments:
\begin{itemize}
  \item[(1)] From Table~\ref{tab:Bstar2P}, it can be seen that the branching fractions of $b\to c \tau \bar{\nu}_{\tau}$ induced $\bar{B}^{*}_{u,d,s}$ decays are at the level of ${\cal O}(10^{-8}-10^{-7})$, while the  $b\to u \tau \bar{\nu}_{\tau}$ induced decays are relatively rare due to the suppression caused by the CKM factor. In addition, the difference between the branching fractions of three decay modes induced by  $b\to c \tau \bar{\nu}_{\tau}$~(or  $b\to u \tau \bar{\nu}_{\tau}$) transition is mainly attributed to the relation of total decay widths, $\Gamma_{\rm tot}(B^{*-}):\Gamma_{\rm tot}(\bar{B}^{*0}):\Gamma_{\rm tot}(\bar{B}^{*0}_{s})\sim 1:2:6$, illustrated by Eqs.~(\ref{eq:GtotBu}), (\ref{eq:GtotBd}) and (\ref{eq:GtotBs}).

  \item[(2)] Comparing with the SM results, one can easily find from Table~\ref{tab:Bstar2P} that ${\cal B}(\bar{B}^*\to P \tau^- \bar{\nu}_{\tau})$ are  enhanced about $20\%$ by the NP contributions of $V_L$ and $V_R$. It is also can be clearly seen from Figs.~\ref{fig:Bstar2PCVob1} (a) and (b). However,  as shown in Figs.~\ref{fig:Bstar2PCVob1} (a) and (b), due to the large theoretical uncertainties caused by the form factors, the NP hints are hard to be totally distinguished from  the SM results.

  \item[(3)] The theoretical uncertainties can be well-controlled  by using the ratio $R^*_P$ instead of decay rate due to the cancellation of  nonperturbative errors, therefore $R^*_P$ is much suitable for probing the NP hints. From the last three rows of Table~\ref{tab:Bstar2P}, it can be found that the NP prediction for $R^*_{P}$ significantly deviates  from the SM result. Especially, as Figs.~\ref{fig:Bstar2PCVob1} (c) and (d) show, the NP effects can be totally distinguished from the SM at $q^2\gtrsim 7\,{\rm GeV^2}$ even though the theoretical errors are considered. So, future measurements on $\bar{B}^*\to P \tau^- \bar{\nu}_{\tau}$ decays  can  make further test on the NP models which provide possible solutions to the $R_{D}$ and $R_{D^*}$ problems.

\item[(4)] From Figs.~\ref{fig:Bstar2PCVob1} (e-h) it can be found that the NP contribution of solution A has little effect on the observables $A_{\lambda}^P$ and  $A_{\theta}^P$ in the whole $q^2$ region, which can be understood from the following analyses. Because the NP contribution of solution A  is dominated by the left-handed coupling  $V_L$, we can find that $|{\cal M}(\bar{B}^{*} \to P \ell^- \bar{\nu}_{\ell})|\propto |(1+V_L)|^2$ in the limit of $(1+V_L)\gg V_R$. As a result,  the NP contributions~(solution A) to the numerator and denominator of $A_{\lambda}^P$ and  $A_{\theta}^P$ cancel each other out to a large extent. For $A_{\lambda}^P$, the cases of solutions B, C and D  are  similar to the solution A.

  \end{itemize}
\subsection{Scenario II: effects of $S_L$ and $S_R$ type couplings}
\begin{figure}[t]
\caption{The allowed spaces of $S_L$ and $S_R$ obtained by fitting to the date of $R_{D}$ and $R_{D^{*}}$. The other captions are the same as in Fig.~\ref{fig:VLVR}.}
\begin{center}
\subfigure[]{\includegraphics[scale=0.5]{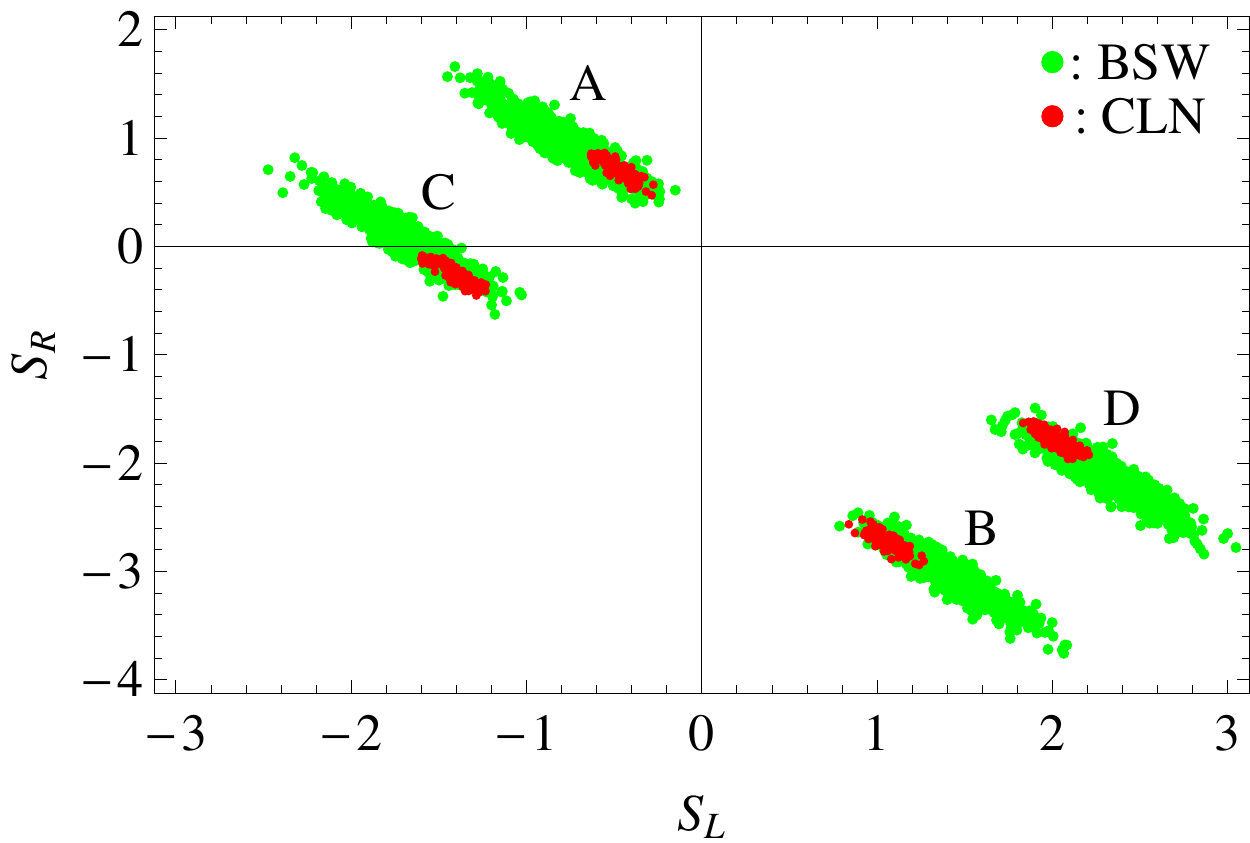}}\qquad\quad
\subfigure[]{\includegraphics[scale=0.5]{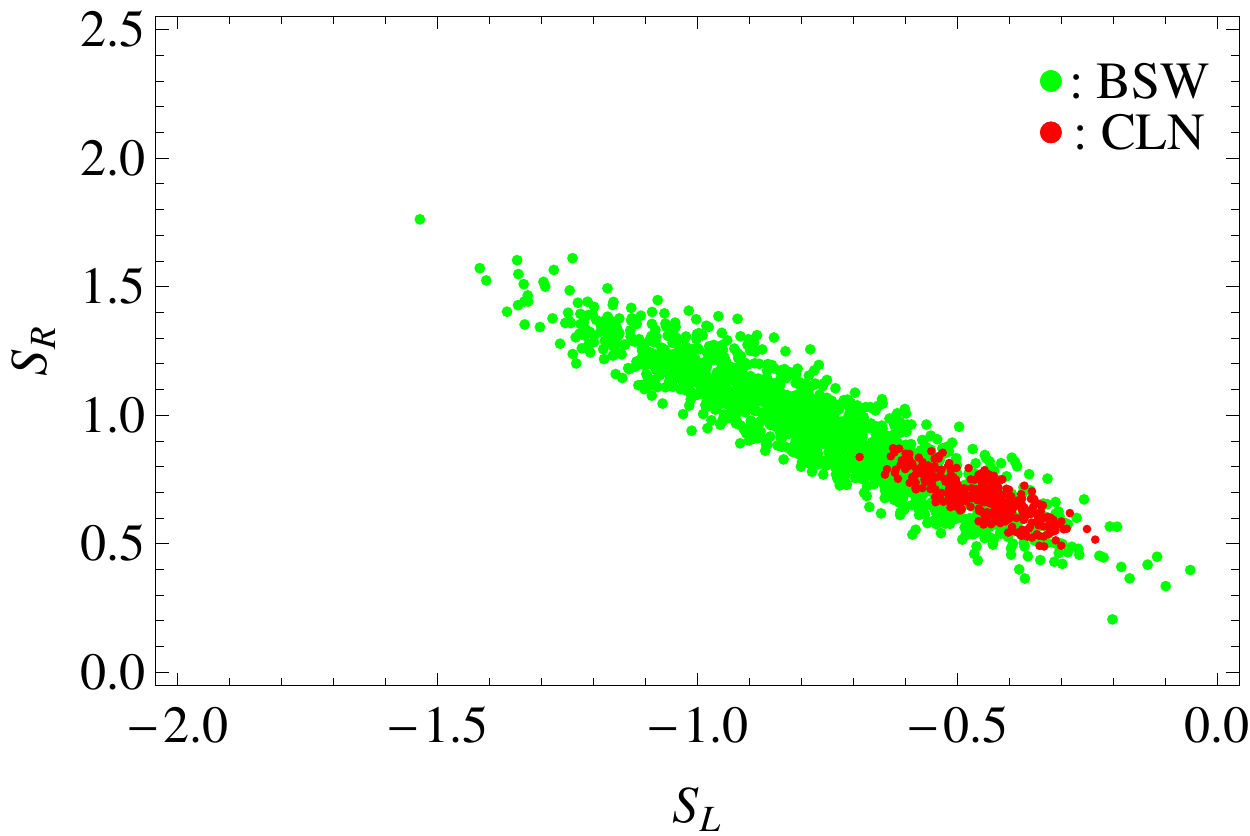}}
\end{center}
\label{fig:SLSR}
\end{figure}
\begin{figure}[t]
\caption{The $q^2$-dependence of the differential observables $d\Gamma/dq^2$, $R^*_P$, $A_{\lambda}^P$ and $A_{\theta}^P$ for $B^{*-} \to D^0 \tau^- \bar{\nu}_{\tau}$ and $\pi^0 \tau^- \bar{\nu}_{\tau}$ decays within the SM and scenario II.}
\begin{center}
\subfigure[]{\includegraphics[scale=0.45]{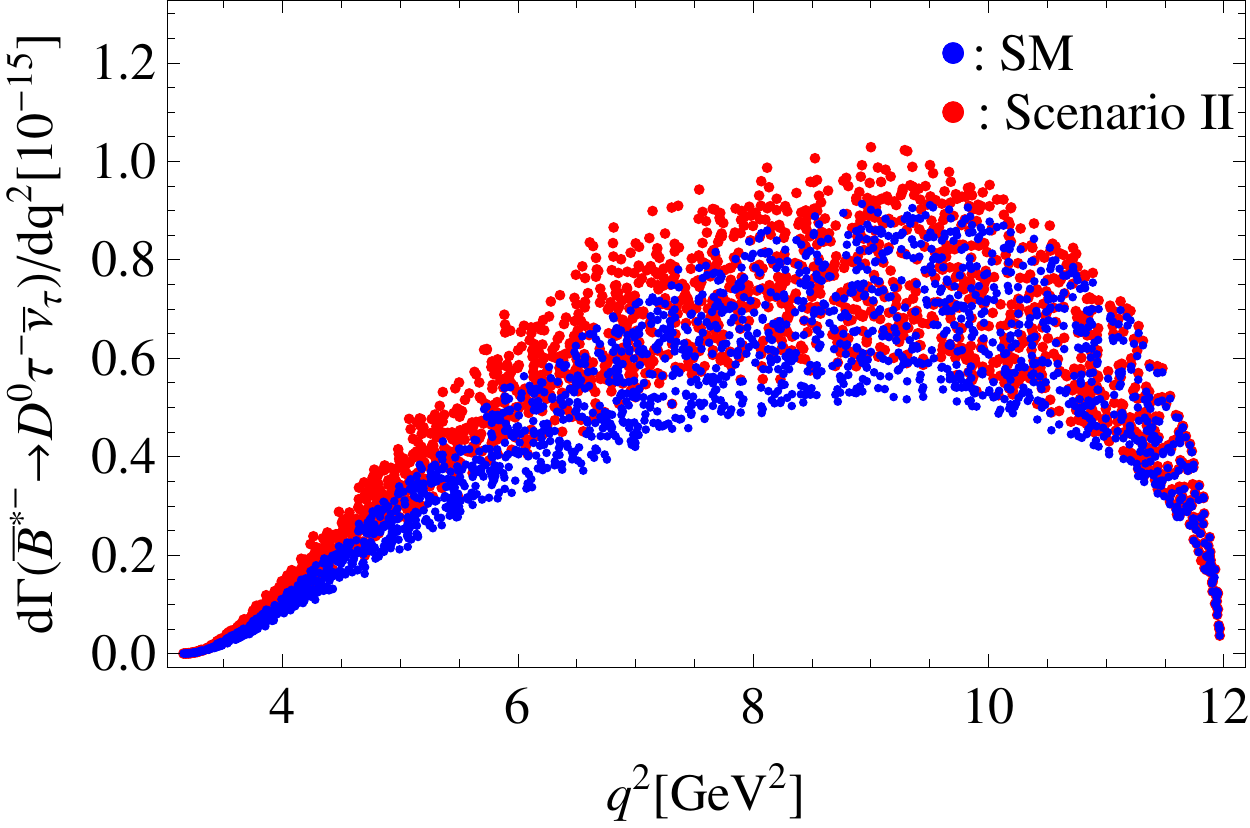}}\qquad\quad
\subfigure[]{\includegraphics[scale=0.45]{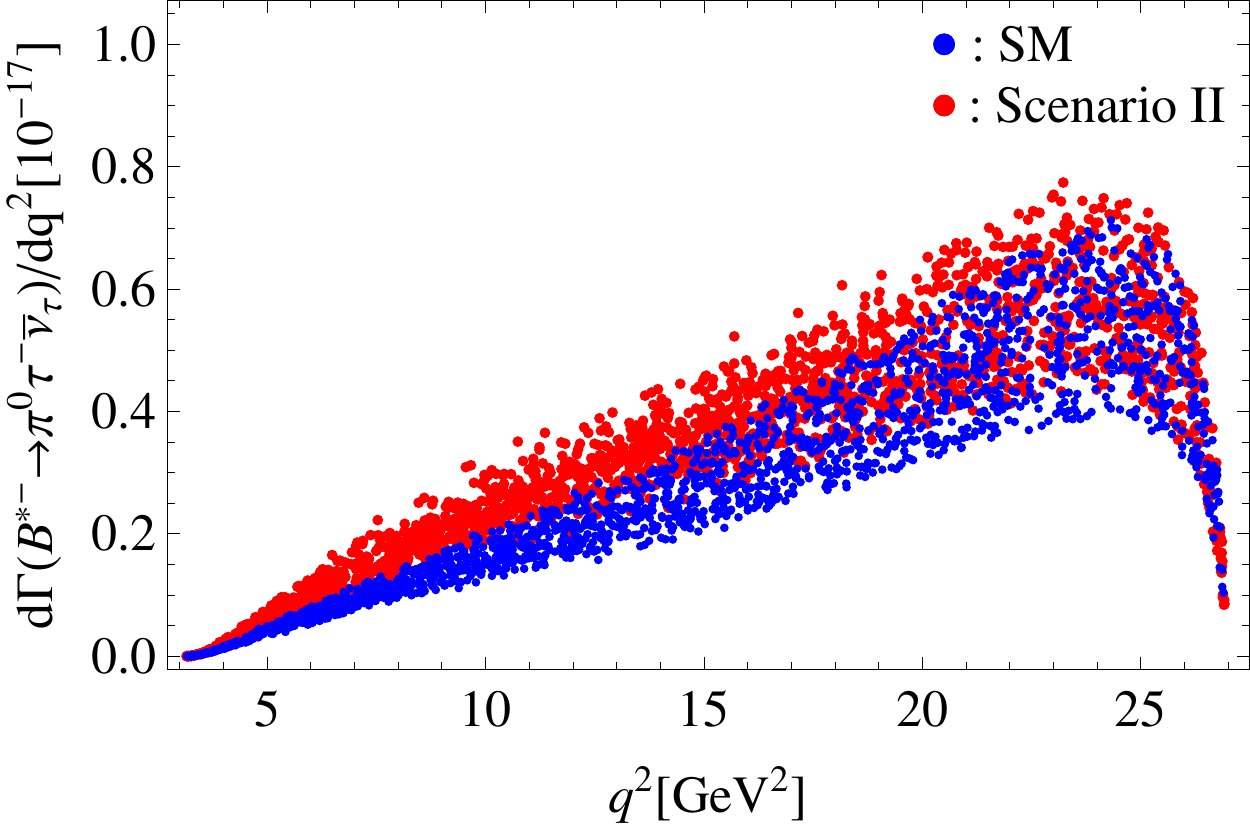}}\\
\subfigure[]{\includegraphics[scale=0.45]{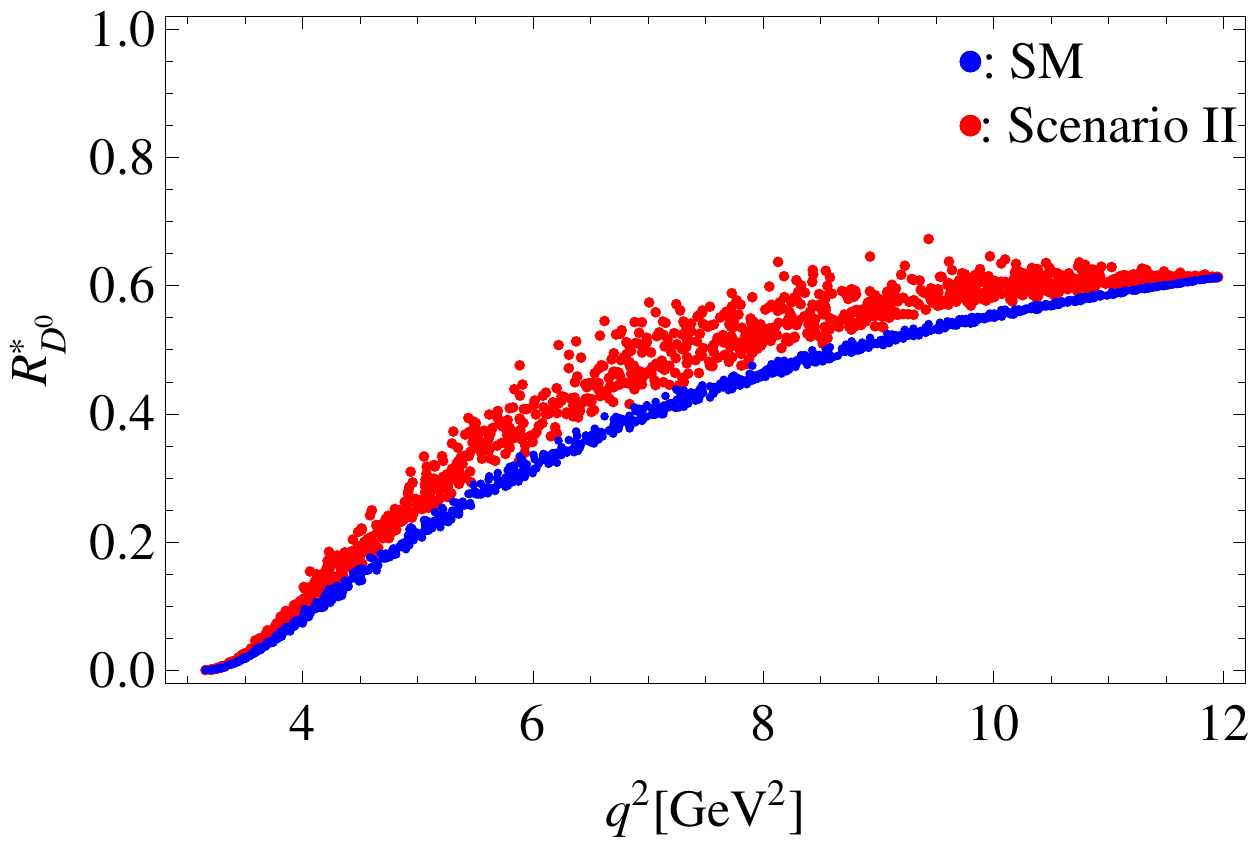}}\qquad\quad
\subfigure[]{\includegraphics[scale=0.45]{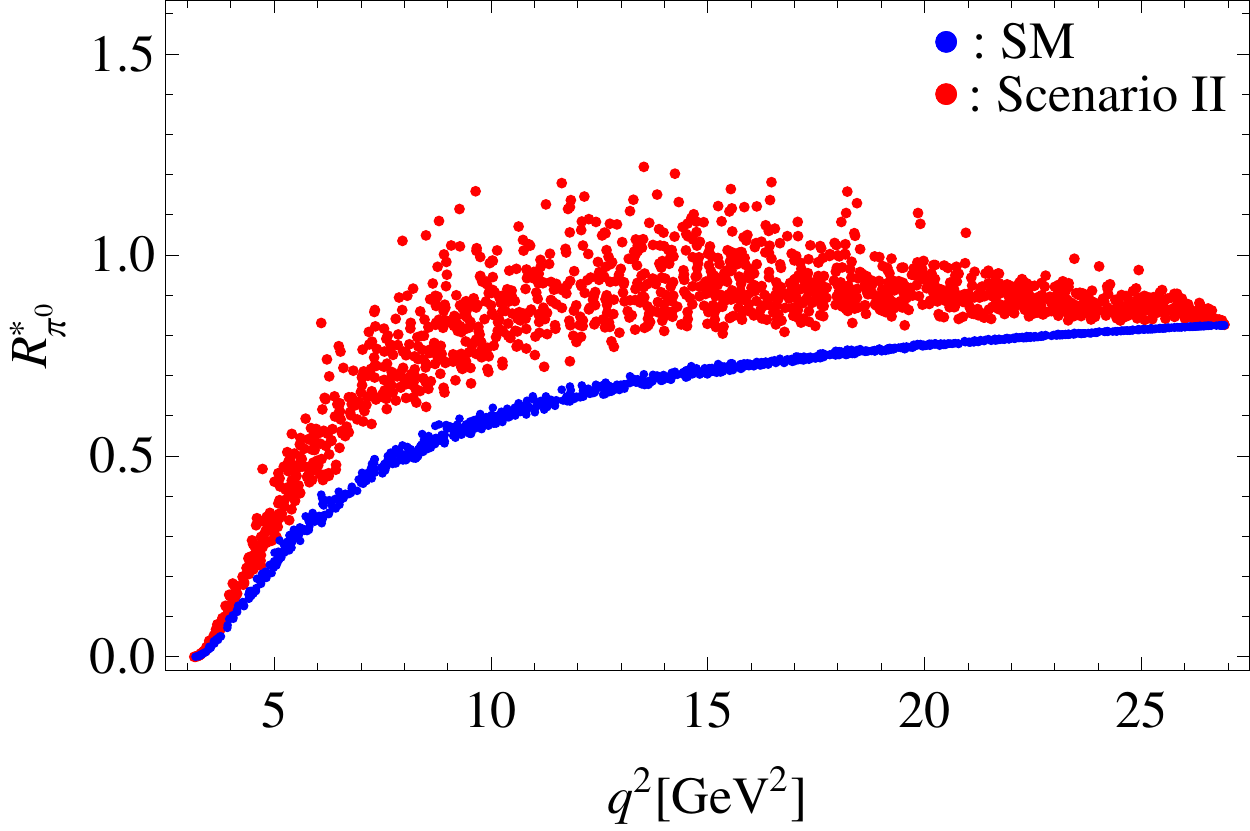}}\\
\subfigure[]{\includegraphics[scale=0.45]{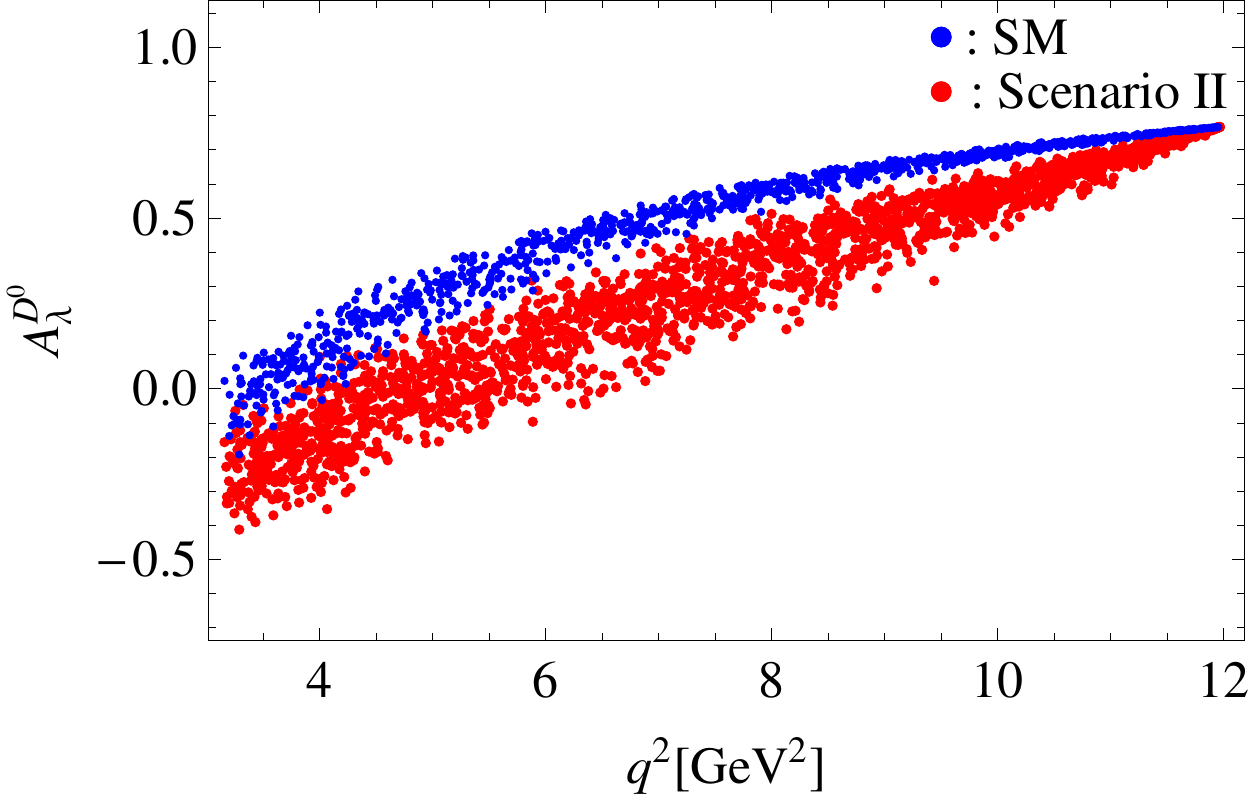}}\qquad\quad
\subfigure[]{\includegraphics[scale=0.45]{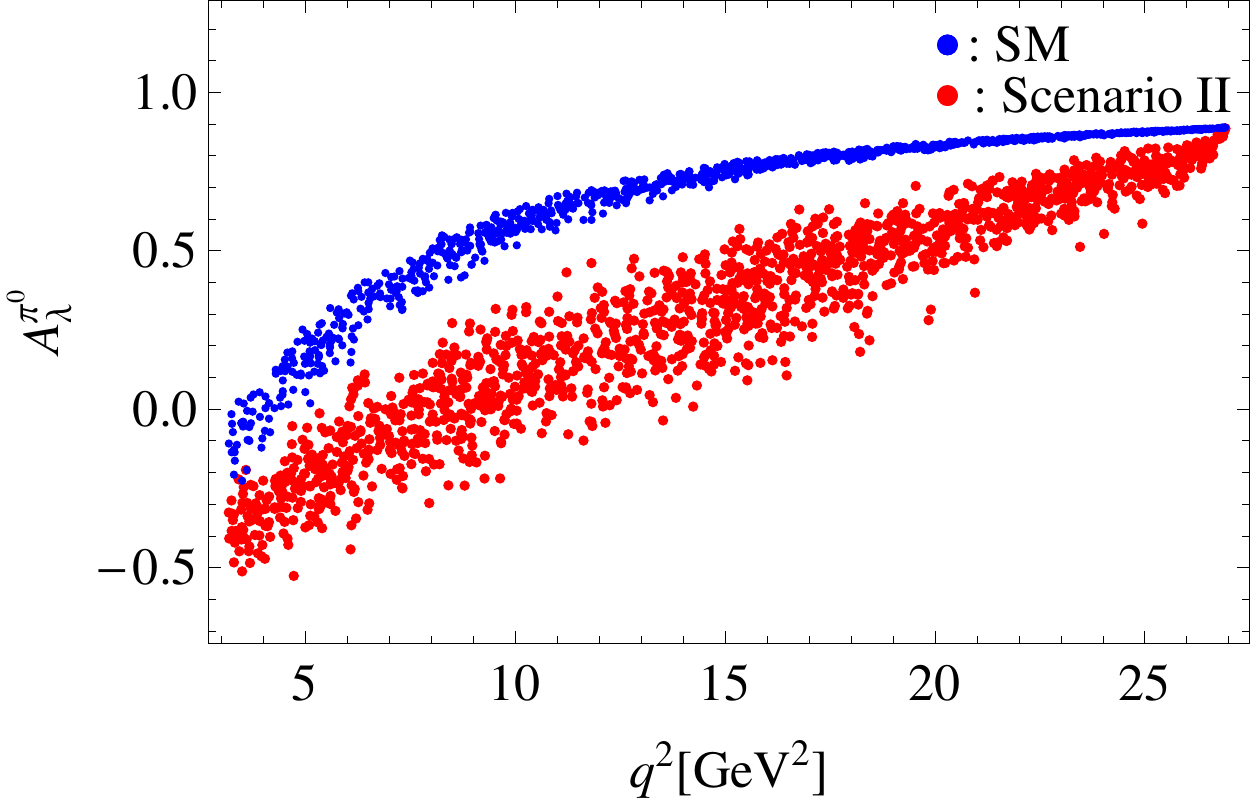}}\\
\subfigure[]{\includegraphics[scale=0.45]{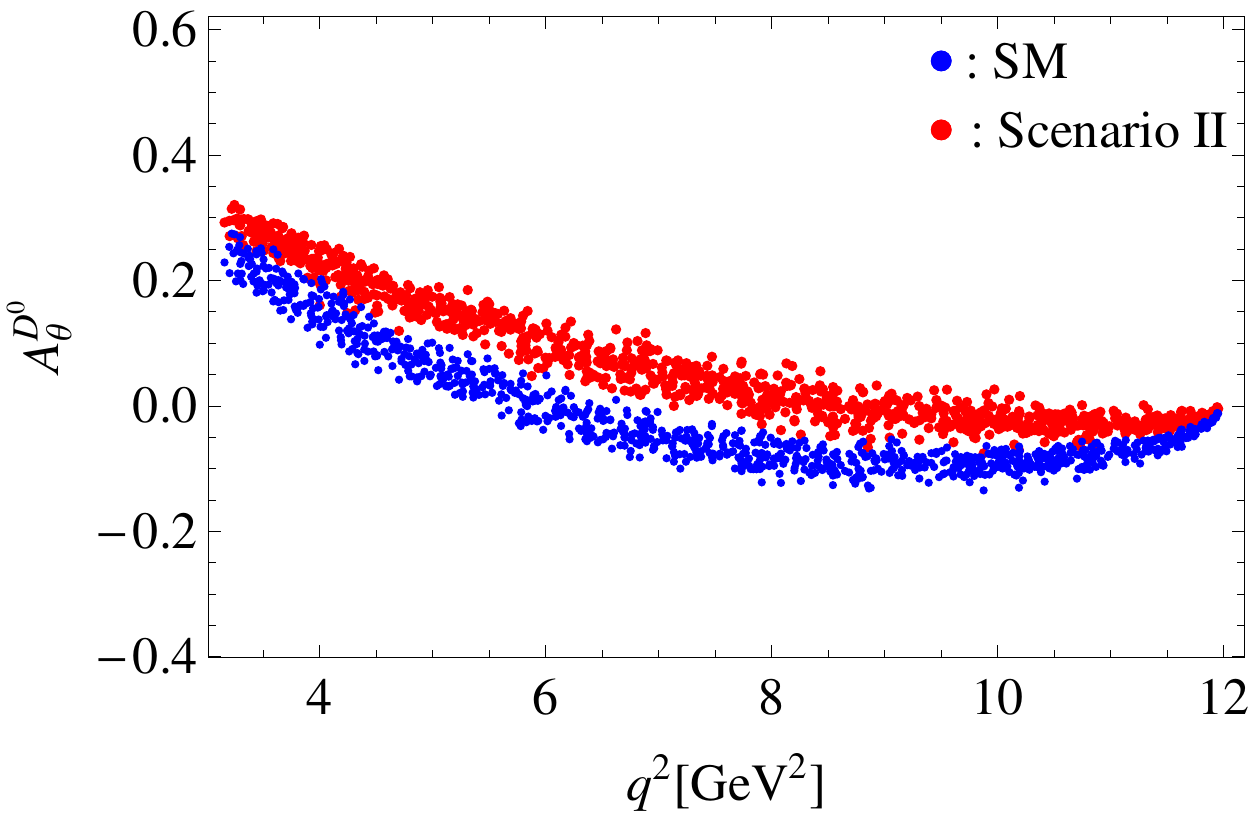}}\qquad\quad
\subfigure[]{\includegraphics[scale=0.45]{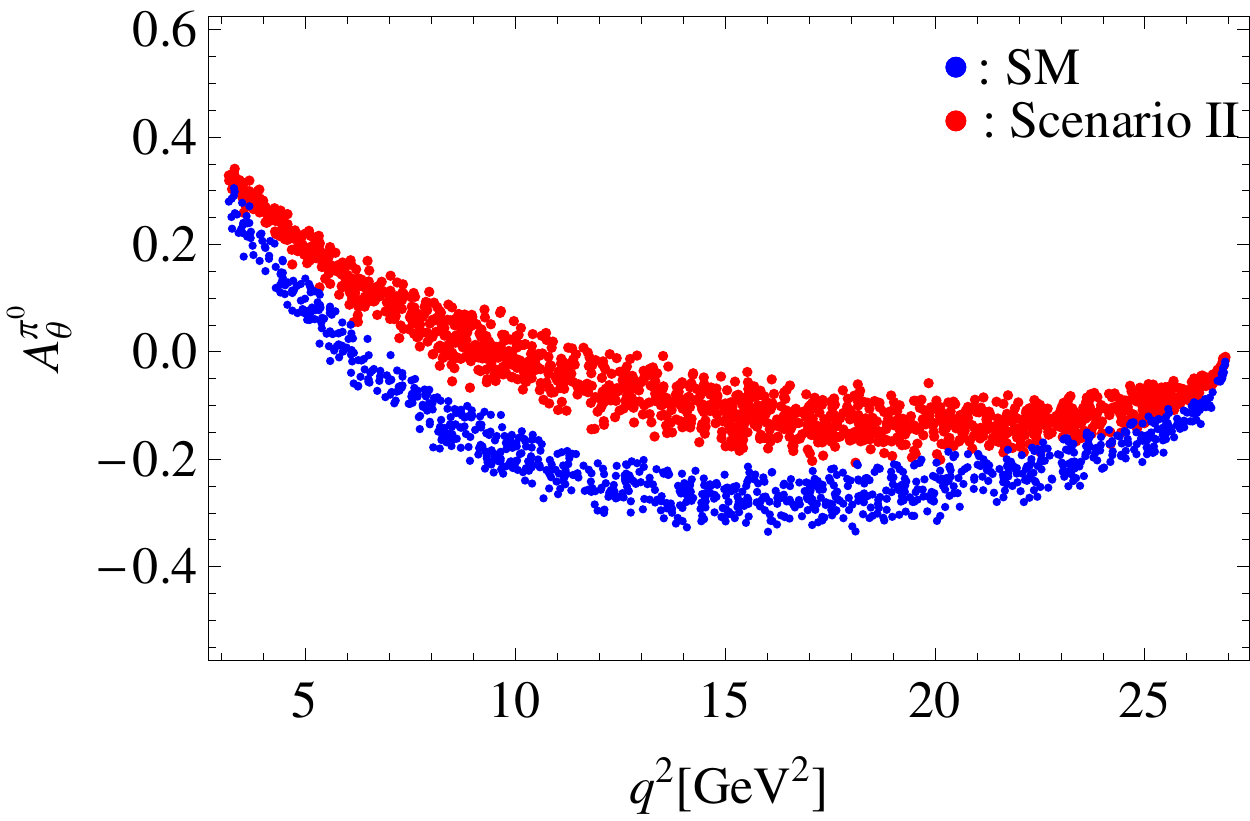}}
\end{center}
\label{fig:Bstar2PCSob1}
\end{figure}

In this subsection, we only consider the effects of scalar interactions $S_L$ and $S_R$ and take the other NP couplings to be zero. Under the $1\sigma$ constraint from the  date of $R_D$ and $R_D^{*}$,  the allowed spaces of $S_L$ and $S_R$ are shown in Fig.~\ref{fig:SLSR}. Similar to the scenario I, four solutions for $S_L$ and $S_R$ are found in scenario II, which can be seen from Fig.~\ref{fig:SLSR}~(a); and the fitting results obtained by using form factors in CLN parametrization and BSW model are in consistence with each other. The solutions B-D result in so large NP contributions; therefore,  in the following discussion, we pay our attention to the solution A, which are replotted in Fig.~\ref{fig:SLSR}~(b). The numerical  result of solution A is
\begin{eqnarray}
S_L=-0.46^{+0.24}_{-0.24}\,,\qquad S_R=0.70^{+0.23}_{-0.24}\,.
\end{eqnarray}
Using these values, we present in Table~\ref{tab:Bstar2P} our numerical predictions of scenario II for the observables, $\mathcal{B}(\bar{B}^*\to P \tau^- \bar{\nu}_{\tau})$ and  $q^2$-integrated $R^*_P$. Moreover, the $q^2$ distributions of differential observables $d\Gamma/dq^2$, $R^*_P$, $A_{\lambda}^P$ and $A_{\theta}^P$ are shown in Fig.~\ref{fig:Bstar2PCSob1}. The following are some discussions for these results:
\begin{itemize}
\item
From Table \ref{tab:Bstar2P} and Figs.~\ref{fig:Bstar2PCSob1}~(a) and (b), it can be found that the ${\cal B}(\bar{B}^*\to P \tau^- \bar{\nu}_{\tau})$ and $R^*_P$ can be  enhanced about $15\%$ compared with the SM results by the NP contributions. Similar to the situation of scenario I, the NP effect of  $S_L$ and $S_R$ on  $R^*_P$  is much significant than the one on branching fraction due to the theoretical uncertainties of $R^*_P$ can be well controlled.  Especially,  as Figs.~\ref{fig:Bstar2PCSob1}~(a) and (b) show, the spectra of the SM and NP for $R^*_P$  can be clearly distinguished at  middle $q^2$ region.
\item
The main difference between the effects of scalar and vector couplings  on the $\bar{B}^*\to P \tau^- \bar{\nu}_{\tau}$ decays is that the former  only contributes to the longitudinal amplitude, which can be found from Eq.~(\ref{eq:SdG}). As a result, their effects on ${\cal B}(\bar{B}^*\to P \tau^- \bar{\nu}_{\tau})$ and $R^*_P$  are  a little different, which can be seen by comparing Figs.~\ref{fig:Bstar2PCVob1}~(a-d) with Figs.~\ref{fig:Bstar2PCSob1}~(a-d).
\item
Another significant difference between the scalar and vector couplings is that only the leptonic helicity amplitudes of scalar type with $\lambda_{\ell}=1/2$ survive, which can be easily found from Eqs.~(\ref{eq:DdGml}) and (\ref{eq:DdGmp}). Therefore, as  Figs.~\ref{fig:Bstar2PCSob1}~(e) and (f) show, the scalar couplings lead to significant  NP effects on the $A^{P}_{\lambda}$, which is obviously different from predictions of  vector couplings  in scenario I (Figs.~\ref{fig:Bstar2PCVob1}~(e) and (f) ). Besides, as Figs.~\ref{fig:Bstar2PCSob1}~(e) and (f) show, $S_L$ and $S_R$ couplings also have large contributions to the  $A^{P}_{\theta}$ at all $q^2$ region, which is another difference with the vector couplings (Figs.~\ref{fig:Bstar2PCVob1}~(g) and (h) ). Therefore, the future measurements on these observables will provide strict tests on the SM and various NP models.
\end{itemize}

\section{Summary}
In this paper, motivated by the observed  ``$R_{D^{*}}$ and $R_{D}$ puzzles" and its implication of NP,  we have studied the NP effects on the $b\to (c\,,u) \ell^- \bar{\nu}_{\ell}$ induced semileptonic $\bar{B}^*_{u,d,s} \to P \ell^- \bar{\nu}_\ell$ ($P=D\,,D_s\,,\pi\,,K$) decays in a model-independent scheme. Using the allowed   spaces of vector and scalar  couplings obtained by fitting to the data of $R_{D^{*}}$ and $R_{D}$, the NP effects on the decay rate,  ratio $R^*_{P}$, lepton spin asymmetry and  forward-backward asymmetry are studied in vector and scalar scenarios respectively. It is found that the vector couplings present large contributions to the decay rate and $R^*_{P}$, but their effects on $A_{\lambda}^P$ and $A_{\theta}^P$ are very tiny. Different from the vector couplings, the scalar couplings present significant effects not only on the  decay rate and $R^*_{P}$ but also on the $A_{\lambda}^P$ and $A_{\theta}^P$. The future measurements on the $\bar{B}^*_{u,d,s} \to P \ell^- \bar{\nu}_\ell$  decays will further test the predictions of the SM and NP, and confirm or refute possible NP solutions to $R_{D^{*}}$ and $R_{D}$.

  \section*{Acknowledgments}
 This work is supported by the National Natural Science Foundation of China (Grant No. 11475055) and the Foundation for the Author of National Excellent Doctoral Dissertation of China (Grant No. 201317).


\end{document}